\documentclass[aps,prl,twocolumn,superscriptaddress,footinbib,reprint]{revtex4-1}
\usepackage{amsmath,amssymb,bm}
\usepackage{graphicx}
\usepackage{epstopdf}
\usepackage{latexsym}
\usepackage{comment}
\usepackage[normalem]{ulem} 
\usepackage[usenames,dvipsnames]{color}
\usepackage[colorlinks]{hyperref}
\hypersetup{
    citecolor=black,
    linkcolor=black,
    filecolor=magenta,      
    urlcolor=black,
    }
\usepackage{natbib}
\usepackage{braket}
\begin{document}

\newcommand{\Rev}[1]{{\color{black}{#1}\normalcolor}} % Revision
\newcommand{\Com}[1]{{\color{red}{#1}\normalcolor}} %Comment

\title{Fast generation of spin squeezing via resonant spin-boson coupling}

\date{\today}
\author{D. Barberena}\thanks{These two authors contributed equally}
\affiliation{JILA, NIST, Department of Physics, University of Colorado,  Boulder, CO 80309, USA}
\affiliation{Center for Theory of Quantum Matter, University of Colorado, Boulder, CO 80309, USA}
\author{S.~R. Muleady}\thanks{These two authors contributed equally}
\affiliation{JILA, NIST, Department of Physics, University of Colorado,  Boulder, CO 80309, USA}
\affiliation{Center for Theory of Quantum Matter, University of Colorado, Boulder, CO 80309, USA}
\author{J.~J. Bollinger}
\affiliation{National Institute of Standards and Technology, Boulder, Colorado 80305, USA}
\author{R.~J. Lewis-Swan}
\affiliation{Homer L. Dodge Department of Physics and Astronomy, The University of Oklahoma, Norman, Oklahoma 73019, USA}
\affiliation{Center for Quantum Research and Technology, The University of Oklahoma, Norman, Oklahoma 73019, USA}
\author{A.~M. Rey}
\affiliation{JILA, NIST, Department of Physics, University of Colorado,  Boulder, CO 80309, USA}
\affiliation{Center for Theory of Quantum Matter, University of Colorado, Boulder, CO 80309, USA}

\begin{abstract}
We propose  protocols for the creation of useful entangled states in a system of spins collectively coupled to a bosonic mode, directly applicable  to trapped-ion and cavity QED setups. The protocols use coherent manipulations of the \Rev{resonant} spin-boson interactions naturally arising in these systems to prepare spin squeezed states exponentially fast in time. \Rev{The resonance condition harnesses the full spin-boson coupling and thus avoids the slower timescales when operating in the off-resonance regime.} We demonstrate the robustness of the protocols by analyzing the effects of natural sources of decoherence in these systems and show their advantage compared to more standard slower approaches where entanglement is generated \Rev{with off-resonant spin-boson interactions}. %algebraically with time. 
\end{abstract}

\maketitle

\noindent{\it Introduction:} 
Spin squeezed states~\cite{Kitagawa1993,Ma2011,Cappellaro_2017,Caves_1980} are a robust example of simple entangled states that can overcome the so called standard quantum limit (SQL) or fundamental noise floor achievable with $N$ uncorrelated particles. Consequently, they have become an important resource for quantum-enhanced sensing and their preparation is a target of intensive research in many different state-of-the-art quantum platforms ~\cite{Cox_2016,Hosten_2016b,Pedrozo-Penafiel2020,Wolfgang_2014,Hosten_2016,Schleier-Smith2010b,Leroux2010c,Bohnet2016,Pezze2018,Qin2019}. While spin squeezing can be generated by diverse mechanisms, a common dynamical approach is one-axis twisting (OAT) \cite{Kitagawa1993}, involving an infinite range Ising interaction between a collection of two level systems. Many schemes to engineer OAT make use of long-range interactions realized in atom-boson platforms, which typically operate in a far-detuned regime where the bosonic degree of freedom just  mediates interactions between the spins \cite{Hu_2017,Pedrozo-Penafiel2020,Bohnet2016}. Consequently, the generated spin-spin interactions are slow compared to the original atom-boson coupling, making any generated squeezing susceptible to decoherence.

Here, we propose to generate spin squeezing through a scheme that \Rev{\emph{resonantly}} couples spins to a bosonic degree of freedom and fully leverages the available atom-boson coupling. This is achieved by implementing 
\Rev{various spin-boson models} that can be concatenated together to prepare a desired final quantum state.\Rev{ We describe a pair of simple protocols to create metrologically useful entanglement exponentially fast, in line with existing proposals and implementations in spin models~\cite{Borregaard_2017,Kitagawa1993,Muessel2015,Hu_2017}, but with the beneficial short timescales associated with the resonant interactions. Furthermore, we characterize the fundamentally achievable phase sensitivity of the protocols  as a function of particle number.} 

In the simplest case, we predict that states with noise variance squeezed by a factor of $\sim N^{-1/2}$ below the SQL can be realized.  We complement the analysis by assessing the performance of the squeezing protocol in the presence of decoherence and determine under what conditions it may outperform standard OAT.  %We also show  Heisenberg limit sensitivity ($\sim N^{-1}$) \cite{Holland_1993,Zurek2001} in principle could be reached by using a time-reversal readout protocol although  at the cost of the system becoming more sensitive to noise in the bosonic degree of freedom.
 
Our results are relevant for a range of platforms that use spin-boson couplings or  pairs of collective  spin ensemble's  for entanglement generation \cite{Norcia2016}. Here, we explicitly demonstrate their utility in $2$D arrays of trapped ions and cavity QED systems~\cite{Bohnet2016}, where previous attempts at spin squeezing using OAT had been  constrained by decoherence. 
\begin{figure*}[t]   
        \centering
        \includegraphics[width=0.98\textwidth]{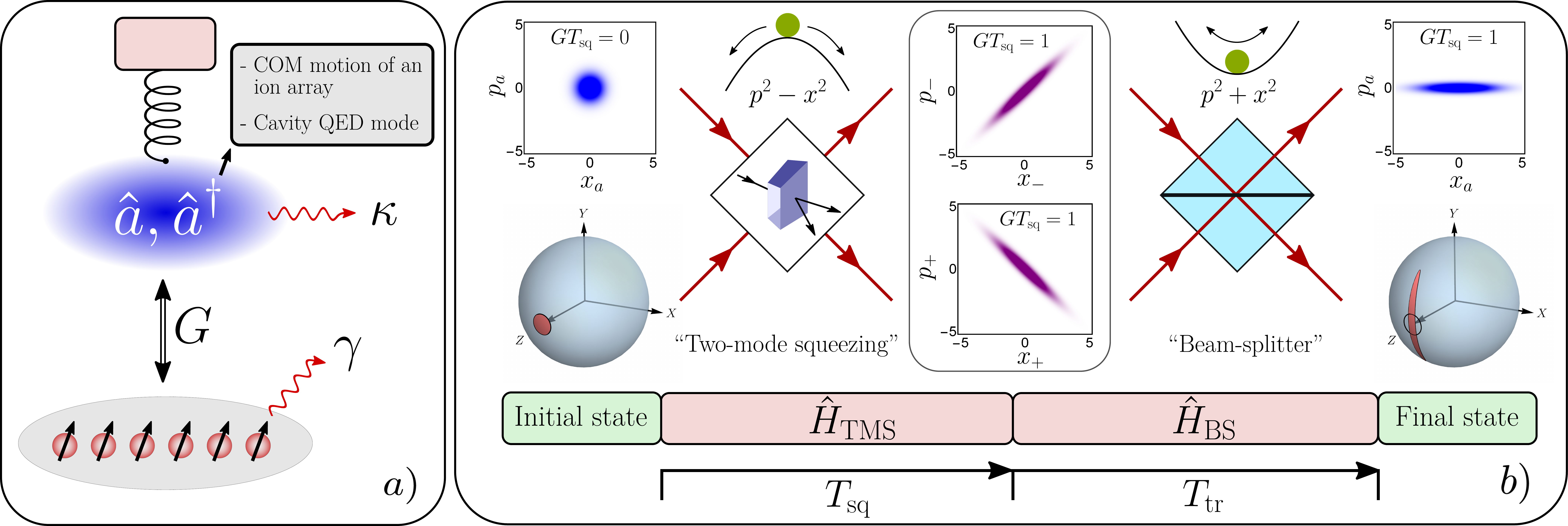}
        \caption{(a) Conceptual schematic of the systems we describe. There is a collection of two-level systems that interacts uniformly with a single harmonic mode. (b) Squeeze and transfer protocol: (i) We begin with a spin state pointing along \Rev{$+z$} and the bosonic mode in vaccuum, represented as phase space distributions. (ii) This initial state is an unstable point of the classical \Rev{Tavis Cummings Hamiltonian [Eq.~\eqref{eq:H_TC}}] and hence exponentially amplifies the quantum fluctuations of the state in a way analogous to spontaneous parametric down conversion (shown in the figure). Near this point, the dynamics are governed by $\hat{H}_{\text{TMS}}$ and create squeezing at $45^\circ$ and $-45^\circ$ in the hybrid $x_+/p_+$ and $x_-/p_-$ planes, respectively. (iii) \Rev{In contrast, for the anti-Tavis-Cummings Hamiltonian [Eq.~\eqref{eq:H_TC}],} the formerly unstable point is now stable and the dynamics are now governed by $\hat{H}_{\text{BS}}$ (represented by the beam-splitter figure). After applying the BS step for a time $T_{\text{tr}}$, the squeezed directions are rotated in the $x_+/p_+$ and $x_-/p_-$ planes and the spin/boson subsystems end up decoupled. In the final state, the squeezed directions align with the $S_x$ and $p_a$ directions.}
        \label{fig:Protocol}
\end{figure*}

\noindent{\it Model:}
\Rev{Our proposal utilizes the Tavis-Cummings (TC)~\cite{Tavis1968} and anti-Tavis-Cummings (ATC) Hamiltonians for $N$ spin-1/2 particles,
\begin{alignat}{3}
        \hat{H}_{\text{TC}}&=-\frac{iG}{\sqrt{N}}(\hat{a}\hat{S}^+-\hat{a}^\dagger\hat{S}^-), \label{eq:H_TC} \\
        \hat{H}_{\text{ATC}}&=-\frac{iG}{\sqrt{N}}(\hat{a}^\dagger\hat{S}^+-\hat{a}\hat{S}^-) , \label{eq:H_ATC}
\end{alignat}
where $\hat{S}_\alpha=1/2\sum_{j=1}^N\hat{\sigma}_j^\alpha$ are collective spin operators for $\alpha=x,y,z$, the $\hat{\sigma}_j^\alpha$ are Pauli matrices acting on the $j$th spin, and $G$ is a coupling constant whose value and $N$ scaling depends on the specific experimental platform. The bosonic degree of freedom is described by the annihilation (creation) operator, $\hat{a}$ ($\hat{a}^\dagger$), while $\hat{S}^\pm$ are spin raising (lowering) operators, traditionally defined as $\hat{S}^{\pm}=\hat{S}_x\pm i\hat{S}_y$.

\Rev{To elucidate the utility of the TC and ATC Hamiltonians for rapid entanglement generation and manipulation, we consider a scenario where} an initial spin state is polarized along \Rev{$+z$} on the collective Bloch sphere. In the large $N$ limit the collective spin operators can be represented by an auxiliary bosonic system with annihilation operator $\hat{b}$, \Rev{using the Holstein-Primakoff~\cite{Holstein1940} relations} $\hat{S}_{\Rev{z}}=N/2-\hat{b}^\dagger\hat{b}$, $\hat{S}^+\approx i\sqrt{N}\hat{b}$, $\hat{S}^-\approx-i\sqrt{N}\hat{b}^\dagger$, such that the coherent spin state pointing along $+z$ corresponds to the bosonic vacuum ($\ket{(N/2)_z}=\ket{0}_b$). The TC and ATC spin-boson models in Eqs.~\eqref{eq:H_ATC} and \eqref{eq:H_TC} then give rise to bosonic two-mode squeezing [$\hat{H}_{\text{TMS}}=\Rev{G}(\hat{a}\hat{b}+\hat{a}^\dagger\hat{b}^\dagger)$] and beam splitter [$\hat{H}_{\text{BS}}=\Rev{G}(\hat{a}^\dagger\hat{b}+\hat{a}\hat{b}^\dagger)$] Hamiltonians, respectively. In terms of the hybrid modes $\hat{c}_{\pm}=(\hat{a}\pm\hat{b})/\sqrt{2}$ and their associated quadratures $\hat{x}_{\pm}=(\hat{c}_{\pm}+\hat{c}^\dagger_{\pm})/\sqrt{2}$, $\hat{p}_{\pm}=(\hat{c}_{\pm}-\hat{c}^\dagger_{\pm})/i\sqrt{2}$, the dynamics created by $\hat{H}_{\text{TMS}}$ and $\hat{H}_{\text{BS}}$ decouple. These modes appear in $\hat{H}_{\text{TMS}}$ in the form of a classically unstable inverted parabolic potential ($p_{\pm}^2-\hat{x}_{\pm}^2$), which amplifies quantum fluctuations in a coherent fashion [see Fig.~\ref{fig:Protocol}(b)]. Conversely, $\hat{H}_{\text{BS}}$ has the form of a stable parabolic potential ($p_{\pm}^2+\hat{x}_{\pm}^2$), which facilitates the perfect transfer of quantum states between spin and boson degrees of freedom. For later convenience, we also define the canonically conjugate quadrature operators $\hat{x}_a=(\hat{a}+\hat{a}^\dagger)/\sqrt{2}$ and $\hat{p}_a=-i(\hat{a}-\hat{a}^\dagger)/\sqrt{2}$.

\noindent{\it Squeeze and transfer protocol  $\rm{ (SnT)}$:} Using the TMS and BS Hamiltonians, we describe a protocol which we refer as SnT (\emph{squeeze and transfer}), that creates squeezing in the spin and boson subsystems simultaneously [see Fig.~\ref{fig:Protocol}(b)]. We begin by (1) preparing the spins polarized along \Rev{$+z$} and the bosons in vacuum $\ket{\psi(0)}=\ket{(N/2)_{\Rev{z}}}\otimes\ket{0}$, and let $\hat{H}_{\text{TC}}$ act so that $\hat{H}_{\text{\text{TMS}}}$ is active. (2) We let the system evolve in this configuration for a time $T_{\text{sq}}$ and then (3) quench $\hat{H}_{\text{TC}}\to -\hat{H}_{\text{ATC}}$ so that now the BS interaction $\hat{H}_{\text{BS}}$ acts for a time $T_{\text{tr}}$. These steps are illustrated in Fig.~\ref{fig:Protocol}(b) using phase space representations of the spin and boson states at different stages of the protocol.

The initial state of the system is shown in Fig.~\ref{fig:Protocol}(b) as two independent isotropic Gaussian distributions in phase space. According to the effective description of $\hat{H}_{\mathrm{TMS}}$, squeezing is created independently in the hybrid modes ($x_+,p_+$) and ($x_-,p_-$). The final BS step, when run for a time $\Rev{G}T_{\text{tr}}=\pi/4$, decouples these hybrid modes such that spins and bosons end up independently squeezed. For this configuration, the resultant spin squeezing is quantified by \cite{Wineland1992}
\begin{equation}\label{eqn:OptimalSpinSqueezing}
    \xi^2\equiv \min_{\theta}N\frac{\braket{(\Delta\hat{S}_\theta)^2}}{\braket{\hat{S}_{\Rev{z}}}^2} \approx e^{-2\Rev{G} T_{\text{sq}}},
\end{equation}
where the variance of the transverse spin $\hat{S}_{\theta} = \cos(\theta)\hat{S}_x + \sin(\theta)\hat{S}_{\Rev{y}}$ is minimized over all angles $\theta$, and all expectation values are calculated at the end of the protocol. When $\xi^2<1$ we say the state is spin squeezed, which also indicates the presence of entanglement~\cite{Ma2011}. The right hand side of Eq.~(\ref{eqn:OptimalSpinSqueezing}) is obtained in the limit of large $N$. Simultaneously, we find that noise in the boson quadrature $\hat{p}_a$ is similarly squeezed by an amount $e^{-2\Rev{G}T_{\text{sq}}}$ \cite{gerry_knight_2004}. \Rev{Under ideal conditions, this exponential growth will only be constrained by curvature effects arising from finite $N$.}} %which generically prevent the squeezing from reaching the Heisenberg limit ($\xi^2_\text{H}=1/N$).} }

\noindent{\it Trapped ion implementation:}
In trapped ion crystals, the TC and ATC models can be indirectly generated by operating at different parameter regimes of the Dicke model~\cite{Dicke1954,Baumann2010,RLS_scrambling_2019,Bollinger_2018,Kirton_2019,Baden2014,Lev2018,Klinder2015}, written in a spin rotated basis as
\begin{equation}\label{eqn:Ham}
    \hat{H}^{\text{ion}}_{\text{Dicke}}=\delta\hat{a}^\dagger\hat{a}+\Omega\hat{S}_x+\frac{2\Rev{g_{\text{ion}}}}{\sqrt{N}}(\hat{a}+\hat{a}^\dagger)\hat{S}_z.
\end{equation}
The spin degree of freedom is encoded in the ion internal levels, while the boson corresponds to a motional mode of the ion array, typically a center of mass mode. The Hamiltonian in Eq.~(\ref{eqn:Ham}) can be realized, for example, in the Penning trap implementation described in Refs.~\cite{Gilmore2020, Bohnet2016,Arghavan_2017}, where the transverse center of mass motional mode of a planar 2D crystal of $N\approx 150$ $\,^9$Be$^+$ ions is coupled to the two $2s\,^{2}S_{1/2}$ valence electron spin states using optical dipole forces (ODF), engineered through a pair of detuned lasers. The electronic states are split by a large magnetic field present in the experiment and can be coherently manipulated using microwaves, which determine the sign and strength of the transverse field $\Omega\hat{S}_x$. The effective oscillator frequency $\delta$ is controlled by the difference between the natural oscillation frequency of the center of mass mode and the detuning between the ODF lasers, while the size and sign of the $N$-independent coupling $g_{\text{ion}}$ is controlled by their intensity and relative phase, respectively. \Rev{The dynamics of this system is then described by Eq.~(\ref{eqn:Ham}) in a frame where the bosons are rotating at the beatnote frequency of the ODF lasers.}

% In trapped ion crystals, the spins are encoded in the ion internal levels, while the collective center of mass motion of the ion array realizes the bosonic degree of freedom. For concreteness of discussion, we consider realizing our proposal in the Penning trap implementation described in Refs.~\cite{Gilmore2020, Bohnet2016,Arghavan_2017}, where the transverse center of mass motional mode of a planar 2D crystal of $N\approx 150$ $\,^9$Be$^+$ ions is coupled to the two $2s\,^{2}S_{1/2}$ valence electron spin states using optical dipole forces (ODF), engineered through a pair of detuned lasers. The electronic states are split by a large magnetic field present in the experiment and can be coherently manipulated using microwaves. In a frame where the bosons are rotating at the beatnote frequency of the ODF lasers, this system is described by the Dicke Hamiltonian~\cite{Dicke1954,Baumann2010,RLS_scrambling_2019,Bollinger_2018,Kirton_2019,Baden2014,Lev2018,Klinder2015}, written in a spin rotated basis as
% \begin{equation}\label{eqn:Ham}
%     \hat{H}^{\text{ion}}_{\text{Dicke}}=\delta\hat{a}^\dagger\hat{a}+\Omega\hat{S}_x+\frac{2g_{\text{ion}}}{\sqrt{N}}(\hat{a}+\hat{a}^\dagger)\hat{S}_z.
% \end{equation}
% The microwave drive determines the strength and sign of the transverse field $\Omega$. The effective oscillator frequency $\delta$ is controlled by the relative detuning of the ODF lasers, while the size and sign of the $N$-independent coupling $g_{\text{ion}}$ is controlled by their intensity and relative phase, respectively.

In the interaction picture induced by $\hat{H}_0=\delta\hat{a}^\dagger\hat{a}+\Omega\hat{S}_x$, the Dicke Hamiltonian Eq.~\eqref{eqn:Ham} takes the form
\begin{equation}\label{eqn:RotatingFrameHamiltonian}
    \hat{H}_{\text{rot}}^{\text{ion}}=-\frac{i\Rev{g_{\text{ion}}}}{\sqrt{N}}(\hat{a}e^{-i\delta t}+\hat{a}^\dagger e^{i\delta t})(\hat{S}_{\text{ion}}^+e^{i\Omega t}-\hat{S}^-_{\text{ion}} e^{-i\Omega t}),
\end{equation}
where $\hat{S}_{\text{ion}}^\pm=\hat{S}_y\pm i\hat{S}_z$ are raising and lowering operators along the $+x$ direction. Assuming $\delta\gg \Rev{g_{\text{ion}}}$ and setting $\delta=\Omega$ or $\delta=-\Omega$ we can then apply a rotating wave approximation to recover Eq.~(\ref{eq:H_TC}) or Eq.~(\ref{eq:H_ATC}), respectively, with $G_{\text{ion}}=\Rev{g_{\text{ion}}}$. Owing to the rotated spin basis, the spins should be initialized along  $+x$ instead of $+z$ for step (1) of the protocol. 

\begin{figure}
        \centering
        \includegraphics[width=0.48\textwidth]{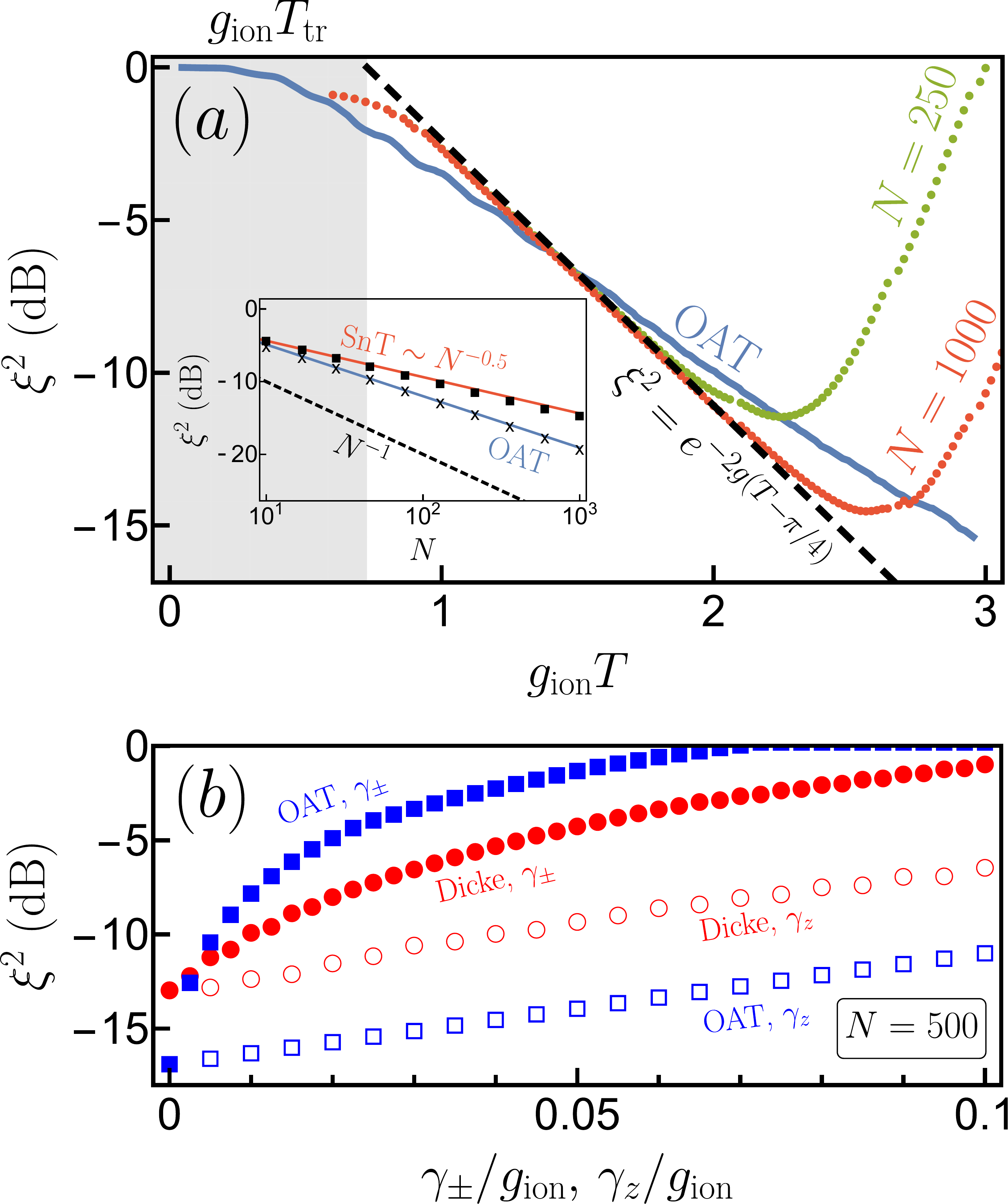}
        \caption{(a) Amount of spin squeezing ($\xi^2$) as a function of total protocol time ($T=T_{\text{sq}}+T_{\text{tr}}$) \Rev{for the ion implementation} \Rev{of the SnT protocol}. The $N\to\infty,\,\delta\to\infty$ result \Rev{is} shown in dashed black and illustrates the exponential improvement of $\xi^2$ with time, starting with an offset corresponding to the transfer step. The dotted lines correspond to the results for finite $\delta=5g_{\Rev{\text{ion}}}$ and $N=250$ (green dots), $N=1000$ (red dots). We also include the results of optimized OAT (as described in the main text) for $N=1000$ in blue. The inset shows the values of the optimal squeezing of both OAT ($\sim N^{-2/3}$, x markers) and \Rev{squeeze and transfer (SnT)} protocols ($\sim N^{-1/2}$, square markers) as a function of $N$ and compare them against the Heisenberg limit ($\xi^2=N^{-1}$). 
        \Rev{(b) SnT optimized over $T_{\text{sq}}$ and $T_{\text{tr}}$ for $N=500$ and $\delta=5g_{\Rev{\text{ion}}}$ as a function of dissipation strength. We consider independently the effects of balanced spin flips ($\gamma_+)$ for both OAT (filled blue squares) and the \Rev{SnT} protocol (filled red circles) and spin dephasing ($\gamma_z$), again for OAT (empty blue squares) and \Rev{SnT} (empty blue circles).}}
        \label{fig:SqueezingIdeal}
\end{figure}

The spin squeezing dynamics given by Eq.~(\ref{eqn:OptimalSpinSqueezing}) for the SnT protocol are only strictly valid in the limit $N\to\infty$, while the trapped ion implementation further requires that $\delta\to \infty$. To benchmark our predictions for finite $N$ and $\delta$, we numerically solve  Eq.~\eqref{eqn:Ham} for $N=250, 1000$ and constraining $\delta=5\Rev{g_{\text{ion}}}$. Given that the result $\Rev{g_{\text{ion}}}T_{\text{tr}}=\pi/4$ technically depends on the previous approximations, in our simulations we optimize over $T_{\text{tr}}$ for each value of $T_{\text{sq}}$ but find that the optimal transfer time remains close to the ideal, with larger deviations occurring at small $\delta$~\cite{SM}. We show this optimized value of $\xi^2$ in Fig.~\ref{fig:SqueezingIdeal}(a) and compare it against Eq.~(\ref{eqn:OptimalSpinSqueezing}). The numerical calculations match the expected exponential squeezing at short times, before slowing and attaining a minimum value due to finite size effects. A systematic analysis of the achievable minimum squeezing with $N$ [see Fig.~\ref{fig:SqueezingIdeal} inset] reveals a scaling $\xi^2\sim N^{-1/2}$ that is reached at $\Rev{g_{\text{ion}}}T_{\text{sq}}\sim (\log N)/4$. \Rev{More sophisticated measurements~\cite{Bohnet2016,Strobel2014,Baamara2021,Holland_1993,Zurek2001}, as well as time-reversal strategies~\cite{Yurke_1986,Hosten_2016,Penasa_2016,Burd2019,Linnemann_2016,Davis_2016,Macri_2016,2021Colombo,SM}, can further improve the sensitivity closer to the Heisenberg limit, though decoherence can impose serious limitations on this~\cite{Huelga1997,Escher2011}.}

For clarity, we compare our results \Rev{for SnT} against the typical squeezing generated by the OAT model in the trapped ion setting
\begin{equation}\label{eqn:OAT}
    \hat{H}_{OAT}=-\frac{4\Rev{g_{\text{ion}}}^2}{\delta N} (\hat {S}_z)^2,
\end{equation} which is realized when $\Omega=0$.  Strictly speaking, OAT only emerges from Eq.~\eqref{eqn:Ham} in the large detuning limit $\delta\gg \Rev{g_{\text{ion}}}$ where the bosons can be adiabatically eliminated. Nevertheless, the OAT dynamics can still be reached in a stroboscopic protocol which only measures the spins at times  $T=2\pi n/\delta$ with $n$ an integer, when they fully decouple from the bosons \cite{Bohnet2016,Wall2017}. We use this approach, numerically solving Eq.~\eqref{eqn:Ham} for $\Omega = 0$, and optimizing the attainable squeezing over $\delta/\Rev{g_{\text{ion}}}$ for each time $T$. For fairness we compare to the total protocol time ($T=T_{\text{tr}}+T_{\text{sq}}$) for a fixed value of $\Rev{g_{\text{ion}}}$ and  $N=1000$. While the overall time duration of our spin-boson protocol initially suffers from the fixed $T_{\text{tr}}$ offset, this is eventually overwhelmed by the exponentially fast TMS dynamics. If we take into account finite size effects, the absolute squeezing attainable with OAT has a more favourable scaling ($N^{-2/3}$ for OAT vs $N^{-1/2}$ for our protocol), but is attained at progressively longer times $\Rev{g^2_{\text{ion}}}T^{\text{opt}}/\delta\sim N^{1/3}$ as the number of particles grows, whereas the time required for our protocol is essentially constant for all experimentally relevant particle numbers ($\Rev{g_{\text{ion}}}T^{\text{opt}}\sim \log N$). 

These shorter time scales can lead to a substantial net gain in spin squeezing in the presence of experimentally relevant decoherence. \Rev{During the preparation stage,} relevant decoherence processes are dephasing and spin flips, with jump operators $\sqrt{\gamma_z}\hat{\sigma}_z^i/2$ and $\sqrt{\gamma_\pm}\hat{\sigma}^{\pm}_i$, respectively. For simplicity of our following discussion we set $\gamma_+=\gamma_-$. In the context of trapped ion simulators such as, e.g., Ref.~\cite{Gilmore2020}, these spin decoherence processes are a consequence of the applied optical dipole force, and bosonic decay is not relevant. 

To efficiently investigate the impact of spin decoherence for moderate size systems we use a semiclassical numerical method \cite{Huber2022}
%, which has been shown to accurately capture the effects of single-particle dissipation for similar spin-boson models, and
which we  benchmark for Eq.~\eqref{eqn:Ham} in \cite{SM}. The results are shown in Fig.~\ref{fig:SqueezingIdeal}(b), where we plot the squeezing of the first protocol as a function of dissipation strength after optimization over $T_{\text{sq}}$ and $T_{\text{tr}}$. We find that SnT is substantially more robust than OAT against spin flips but less robust against dephasing. In the latter case, a large fraction of the discrepancy can be accounted for by finite size effects. Using Ref.~\cite{Gilmore2020} as a state-of-the-art example, typical decoherence rates are $\gamma_z=610\text{ s}^{-1}\approx 0.05g_{\Rev{\text{ion}}}$ and $\gamma_+=61\text{ s}^{-1}\approx 0.005g_{\Rev{\text{ion}}}$ ($g_{\Rev{\text{ion}}} = 2\pi \times 2$~kHz, $\Omega\sim 2\pi\times 10-15$~kHz). For this set of parameters, Fig.~\ref{fig:SqueezingIdeal} indicates that dephasing is the main limiting factor, with an expected squeezing of about $9$ dB.

% In cavity QED systems, the interaction between two-level atoms (spin subsystem) and an electromagnetic cavity mode (boson subsystem) is naturally described by the Tavis-Cummings model [Eq.~(\ref{eq:H_TC})] in the rotating frame of the atoms when the cavity is tuned on resonance with the atomic transition. As an example, we consider the cavity implementation discussed in \cite{Norcia2016}, where $N\sim 10^5$ $^{88}\text{Sr}$ atoms are trapped and cooled in a 1D optical lattice, and a high-finesse optical cavity is collectively coupled to the corresponding $^1S_0\to\,^3P_1$ optical transition. An important feature of these cavity systems is that the effective coupling constant $G$ is $N$-dependent, i.e. $G_\text{cav}= g_\text{cav}\sqrt{N}$, where $2g_{\text{cav}}$ is the $N$-independent single photon Rabi frequency. The resulting atom-light coupling is described by the Hamiltonian
% \begin{alignat}{3}
%         \hat{H}_{\text{TC}}^{\text{cav}}&= \Delta_{\text{cav}}\hat{a}^\dagger\hat{a} -ig_{\text{cav}}(\hat{a}\hat{S}^+-\hat{a}^\dagger\hat{S}^-) \label{eq:H_TC_cav}
% \end{alignat}
% in the rotating frame of the atoms, where $\Delta_{\text{cav}}$ describes the atom-cavity detuning.

\noindent{\it Cavity QED implementation:} \Rev{In cavity QED systems, the TC model [Eq.~(\ref{eq:H_TC})] arises as the natural interaction between two-level atoms and an electromagnetic cavity mode. In the rotating frame of the atoms, the coupling between these degrees of freedom is described by
\begin{alignat}{3}
        \hat{H}_{\text{TC}}^{\text{cav}}&= -\Delta_{\text{cav}}\hat{a}^\dagger\hat{a} -ig_{\text{cav}}(\hat{a}\hat{S}^+-\hat{a}^\dagger\hat{S}^-) \label{eq:H_TC_cav}
\end{alignat}
where $\Delta_{\text{cav}}$ is the detuning of the cavity with respect to the atomic transition frequency, $2g_{\text{cav}}$ is the single photon Rabi frequency, and $\hat{H}_{\text{TC}}$ [Eq.~(\ref{eq:H_TC})] is recovered when $\Delta_{\text{cav}}\to 0$. A direct comparison between Eq.~(\ref{eq:H_TC_cav}) and Eq.~(\ref{eq:H_TC}) highlights a very important feature of cavity systems: the effective coupling constant $G$ is $N$-dependent, i.e. $G_\text{cav}= g_\text{cav}\sqrt{N}$. For definiteness, we now consider the cavity implementation discussed in \cite{Norcia2016,Muniz2020}, where $^{88}\text{Sr}$ atoms are trapped and cooled in a 1D optical lattice, and a high-finesse optical cavity is collectively coupled to the $^1S_0\to\,^3P_1$ optical transition.

For the SnT protocol we set $\Delta_{\text{cav}} = 0$. The totally inverted initial spin state along $+z$ [see Fig.~\ref{fig:Protocol}(b)] can be prepared by optical pumping and then steps (1) \& (2) of the squeeze and transfer protocol proceed automatically. To perform the quench $\hat{H}_{\text{TC}}\to\hat{H}_{\text{ATC}}$, a rapid $\pi$ pulse along $x$ can be applied just before step (3), so that $\hat{S}^\pm\to\hat{S}^\mp$. Details including the effects of finite pulse duration are addressed in \cite{SM}, but do not pose a practical limitation for our scheme. We also note that no external drive is present during the entangling stage of the protocol [step (1)], in a similar spirit to other proposals~\cite{Hu_2017}.}

% In cavity-QED platforms the leakage of photons through the cavity mirrors is of key relevance \cite{Baumann2010,Kirton_2019,Klinder2015,Baden2014,Lev2018} and is typically modelled with jump operator $\sqrt{\kappa}\hat{a}$. \Rev{We also consider the effects of spontaneous emission with jump operators $\sqrt{\gamma}\hat{\sigma}^-_j$, and define the single-atom cooperativity $C = (2g_{\text{cav}})^2/(\kappa \gamma)$...}

\begin{figure}
    \centering
    \includegraphics[width=0.48\textwidth]{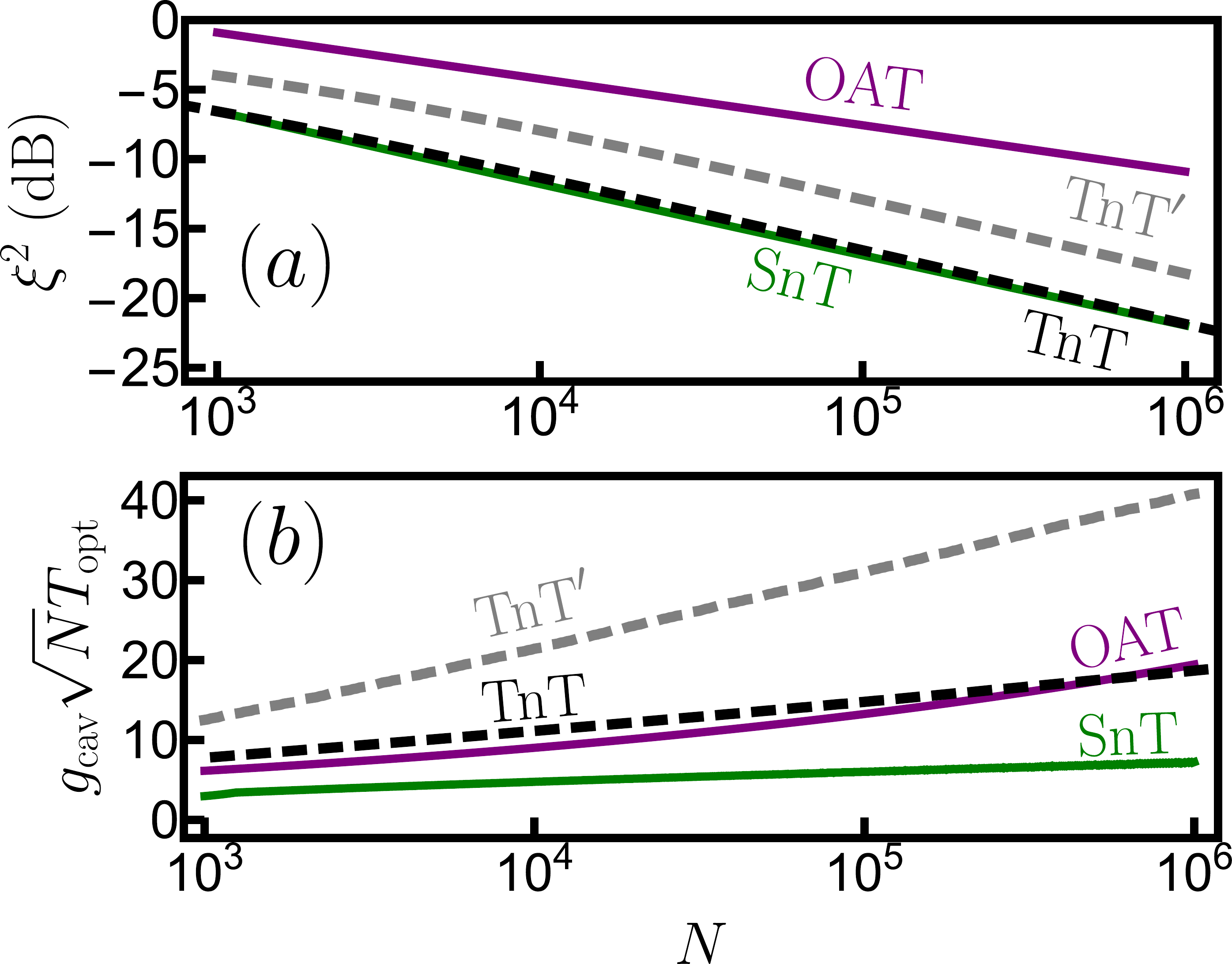}
    \caption{\Rev{(a) Optimal squeezing for various protocols: squeeze and transfer (SnT, solid green), one axis twisting (OAT, solid purple), twist and turn (TnT, dashed black) as a function of $N$ for fixed $(g_{\text{cav}},\gamma,\kappa)=2\pi\times (10.9,7.5,153) \text{ kHz}$. We also include a modified curve for TnT that takes into account its regime of validity ($\text{TnT}'$, dashed gray). (b) Optimal evolution time for the same set of protocols.}}
    \label{fig:Dissipation}
\end{figure}

Relevant decoherence processes affecting the protocol come from leakage of photons through the cavity mirrors~\cite{Baumann2010,Kirton_2019,Klinder2015,Baden2014,Lev2018}, modelled with jump operator $\sqrt{\kappa}\hat{a}$, and spontaneous emission, with jump operator $\sqrt{\gamma}\hat{\sigma}^-_j$. To quantify these effects, we numerically solve the associated Heisenberg equations of motion in the limit of large $N$~\cite{SM} for fixed $\kappa/g_{\text{cav}}$, $\gamma/g_{\text{cav}}$, and compute the squeezing parameter as a function of $N$ after optimization over $T_{\text{tr}}$ and $T_{\text{sq}}$. We use the values $(g_{\text{cav}},\gamma,\kappa)=2\pi\times (10.9,7.5,153) \text{ kHz}$~\cite{Norcia2016,Muniz2020}, and vary $N$ in the range $(10^3,10^6)$. The results are shown in Fig.~\ref{fig:Dissipation}(a), which indicates that the optimal squeezing scales as $N^{-1/2}$.

\Rev{We also compare SnT against OAT and twist-and-turn (TnT)~\cite{Muessel2015} protocols, both of which can be engineered by operating the cavity in a far-detuned regime $|\Delta_{\text{cav}}|\gg g_{\text{cav}}\sqrt{N}$~\cite{Hu_2017}:
\begin{align}\label{eqn:TnTHamiltonian}
    \hat{H}_{\text{OAT}}^{\text{cav}}=\chi_{\text{cav}}\hat{S}_z^2,\hspace{1cm} \hat{H}_{\text{TnT}}^{\text{cav}}=\chi_{\text{cav}}(\hat{S}_z^2+\hat{S}_x),
\end{align}
where $\chi_{\text{cav}}=4g^2\Delta_{\text{cav}}/(4\Delta_{\text{cav}}^2+\kappa^2)$. In the presence of the same decoherence sources, optimization of squeezing over $\Delta_{\text{cav}}/\kappa$ and evolution time leads to $\xi^2_{\text{OAT}}=6(C N)^{-1/3}$~\cite{Hu_2017,LewisSwan2018} and $\xi^2_{\text{TnT}}\approx 4.6(C N)^{-1/2}$~\cite{Hu_2017,SM}, where $C=4g^2_{\text{cav}}/(\kappa\gamma)$ is the cooperativity parameter. These results are also shown in Fig.~\ref{fig:Dissipation}(a) as a function of $N$ for the same values of $g_{\text{cav}},\gamma,\kappa$ as for SnT. However, with these set of parameters the optimal detuning for TnT satisfies $\Delta_{\text{cav}}^{\text{opt}}\approx 2.5 g_{\text{cav}}\sqrt{N}$, which is outside the regime of validity of $\hat{H}_{\text{TnT}}^{\text{cav}}$ in Eq.~(\ref{eqn:TnTHamiltonian}). Fixing instead $\Delta_{\text{cav}}=10 g_{\text{cav}}\sqrt{N}$ to guarantee the applicability of Eq.~(\ref{eqn:TnTHamiltonian}) and optimizing over evolution time leads to the gray line labelled $\text{TnT}'$ in Fig.~\ref{fig:Dissipation}. For completeness, in Fig.~\ref{fig:Dissipation}(b) we show the total protocol time, measured in units of $(g_{\text{cav}}\sqrt{N})^{-1}$, as a function of $N$.}

It is possible to obtain analytical formulas when $\kappa\gg \gamma$. Then the achievable spin squeezing is~\cite{SM},
\begin{equation}
    \xi^2_{\text{opt}}= \frac{ \pi \kappa}{8\Rev{g_{\text{cav}}\sqrt{N}}}\approx \frac{ 0.4 \kappa}{\Rev{g_{\text{cav}}\sqrt{N}}},
\end{equation}
obtained at $\Rev{g_{\text{cav}}\sqrt{N}}T_{\text{tr}}=\pi/4+\kappa/(4g_{\text{cav}}\sqrt{N})$ and $2g_{\text{cav}}\sqrt{N}T_{\text{sq}}\gtrsim \log\big(2.5g_{\text{cav}}\sqrt{N}/\kappa\big)$. This makes manifest the $N^{-1/2}$ scaling of the optimal squeezing and optimal protocol time (up to logarithmic corrections). For $N\approx 10^5$ particles, 15 dB of squeezing are attainable.

%  Using the $^1S_0\to\,^3P_1$ transition in $^{88}\text{Sr}$ and assuming $N=10^5$ atoms, relevant experimental parameters are $g_{\text{cav}}=2\pi\times 11 \text{ kHz}$, $\gamma_z=0$, $\gamma_{-}=2\pi\times 7.5\text{ kHz}=0.002\Rev{g_{\text{cav}}\sqrt{N}}$ and $\kappa=2\pi\times 160\text{ kHz}\approx 0.05\Rev{g_{\text{cav}}\sqrt{N}}$  \cite{Norcia2016}. This leads to an expected squeezing of about $15$ dB [see Fig.~\ref{fig:Dissipation}, lower panel].

% \Rev{To compare our scheme to typical off-resonant protocols, such as \cite{LewisSwan2018} ...}

\noindent{\it Outlook}: We have discussed a protocol that takes full advantage of the on resonance spin-boson coupling and leads to exponentially fast generation of squeezing mitigating  the impact of decoherence. The protocol is relevant for a range of spin-boson platforms at the cutting edge for quantum sensing applications. Although so far we study only deterministic protocols, an interesting direction would be to investigate how dynamical evolution combined with non-demolition measurements improves their performance. 

\begin{acknowledgments}
\emph{Acknowledgments --} 
We thank K. Lehnert  and J.~K. Thompson  for a careful reading and comments on the manuscript, and D. J. Young and E. Y. Song for discussions of experimental implementations. This work is supported by the AFOSR grant FA9550-18-1-0319,  by the DARPA and ARO grant W911NF-16-1-0576, the ARO award W911NF-19-1-0210, NSF JILA-PFC PHY-1734006 and NSF PHY-1820885, QLCI-OMA -2016244, by the U.S. Department of Energy, Office of Science, National Quantum Information Science Research Centers Quantum Systems Accelerator, and by NIST. R.~J. L-S acknowledges support by NSF Grant No. PHY-2110052.
\end{acknowledgments}

\bibliography{library}

\end{document}

\newcommand{\Rev}[1]{{\color{black}{#1}\normalcolor}} % Revision
\newcommand{\Com}[1]{{\color{red}{#1}\normalcolor}} %Comment

\renewcommand{\theequation}{S\arabic{equation}}
\renewcommand{\thefigure}{S\arabic{figure}}

\title{Supplemental Material: Fast generation of spin squeezing via resonant spin-boson coupling}
\date{\today}

\date{\today}
\author{D. Barberena}
\affiliation{JILA, NIST, Department of Physics, University of Colorado,  Boulder, CO 80309, USA}
\affiliation{Center for Theory of Quantum Matter, University of Colorado, Boulder, CO 80309, USA}
\author{S.~R. Muleady}
\affiliation{JILA, NIST, Department of Physics, University of Colorado,  Boulder, CO 80309, USA}
\affiliation{Center for Theory of Quantum Matter, University of Colorado, Boulder, CO 80309, USA}
\author{J.~J. Bollinger}
\affiliation{National Institute of Standards and Technology, Boulder, Colorado 80305, USA}
\author{R.~J. Lewis-Swan}
\affiliation{Homer L. Dodge Department of Physics and Astronomy, The University of Oklahoma, Norman, Oklahoma 73019, USA}
\affiliation{Center for Quantum Research and Technology, The University of Oklahoma, Norman, Oklahoma 73019, USA}
\author{A.~M. Rey}
\affiliation{JILA, NIST, Department of Physics, University of Colorado,  Boulder, CO 80309, USA}
\affiliation{Center for Theory of Quantum Matter, University of Colorado, Boulder, CO 80309, USA}

\maketitle  

\section{Squeezing under the ideal protocol}
As in the main text, our starting point are the 
% we begin with the Dicke Hamiltonian,
% \begin{equation}
%     \hat{H}=\delta\hat{a}^\dagger\hat{a}+\Omega\hat{S}_x+\frac{2g}{\sqrt{N}}\hat{S}_z(\hat{a}+\hat{a}^\dagger),\label{eq:Ham_supp}
% \end{equation} 
% where $\hat{a}$ ($\hat{a}^\dagger$) are annihilation (creation) operators describing a bosonic degree of freedom and $\hat{S}_\alpha=1/2\sum_{j=1}^N\hat\sigma_j^\alpha$ are collective spin operators for $\alpha=x,y,z$, and $\sigma_j^\alpha$ are the Pauli matrices acting on the $j$th spin. In the interaction picture induced by $\hat{H}_0=\delta\hat{a}^\dagger\hat{a}+\Omega\hat{S}_x$, the interaction Hamiltonian is
% \begin{equation}
%     \hat{H}_{\text{int}}=-\frac{ig}{\sqrt{N}}(\hat{a}e^{-i\delta t}+\hat{a}^\dagger e^{i\delta t})(\hat{S}^+e^{i\Omega t}-\hat{S}^-e^{-i\delta t}),
% \end{equation}
% where $\hat{S}^\pm\equiv\hat{S}_y\pm i\hat{S}_z$ are spin raising and lowering operators along the $x$ direction. Within the rotating wave approximation, this gives rise to the 
Tavis-Cummings and anti-Tavis-Cummings models, respectively:
\begin{alignat}{3}
       \hat{H}_{\text{TC}}&=-\frac{iG}{\sqrt{N}}(\hat{a}\hat{S}^+-\hat{a}^\dagger\hat{S}^-),\label{eq:H_TC}\\
       \hat{H}_{\text{ATC}}&=-\frac{iG}{\sqrt{N}}(\hat{a}^\dagger\hat{S}^+-\hat{a}\hat{S}^-). \label{eq:H_ATC}
        %\nonumber        \hat{H}_{rot}&=-\frac{4g^2}{\delta N}(\hat{S}_z)^2\quad &&\text{for}\quad g\ll\delta,\Omega=0,
\end{alignat}
When $N$ is large, we can approximately describe the spin degree of freedom using a bosonic mode $\hat{b}$ such that $\hat{S}^+\approx i\sqrt{N}\hat{b}$, $\hat{S}^-\approx-i\sqrt{N}\hat{b}^\dagger$, $\hat{S}_z\approx \frac{N}{2}$, and $\ket{N/2_z}=\ket{0}_b$ (the quantum state with spin pointing along $+z$ is the bosonic vacuum). Under these conditions, we obtain the two-mode squeezing and beam-splitter models known from quantum optics:
\begin{align}\begin{split}\label{eqn:SuppIdealTMSandBS}
        \hat{H}_{\text{TMS}}&=G(\hat{a}^\dagger\hat{b}^\dagger+\hat{a}\hat{b}) ,\\
        \hat{H}_{\text{BS}}&=G(\hat{a}\hat{b}^\dagger+\hat{a}^\dagger\hat{b}) .
        %\nonumber        \hat{H}_{rot}&=-\frac{4g^2}{\delta N}(\hat{S}_z)^2\quad &&\text{for}\quad g\ll\delta,\Omega=0,
\end{split}\end{align}
The squeezing protocol described in the main text is then mathematically represented as 
\begin{equation}
    \ket{\psi_f}=e^{-i\hat{H}_{\text{BS}}T_{\text{tr}}}e^{-i\hat{H}_{\text{TMS}}T_{\text{sq}}}\ket{0}_a\otimes\ket{0}_b,
\end{equation}
where $\ket{0}_a$ is the vacuum of the original boson degree of freedom. In this ideal case, we set the transfer time to $GT_{\text{tr}}=\pi/4$, which generates the following transformation of the boson variables:
\begin{align}\begin{split}
    &e^{-i\hat{H}_{\text{BS}}T_{\text{tr}}}\hat{a}\,e^{i\hat{H}_{\text{BS}}T_{\text{tr}}}=\frac{\hat{a}-i\hat{b}}{\sqrt{2}} ,\\
    &e^{-i\hat{H}_{\text{BS}}T_{\text{tr}}}\hat{b}\,e^{i\hat{H}_{\text{BS}}T_{\text{tr}}}=\frac{\hat{b}-i\hat{a}}{\sqrt{2}} .
\end{split}\end{align}
This induces the following transformation of the $\hat{H}_{\text{TMS}}$:
\begin{equation}
    e^{-i\hat{H}_{\text{BS}}T_{\text{tr}}}\hat{H}_{\text{TMS}}\,e^{i\hat{H}_{\text{BS}}T_{\text{tr}}}=iG\bigg[\frac{(\hat{b}^\dagger)^2-\hat{b}^2}{2}+\frac{(\hat{a}^\dagger)^2-\hat{a}^2}{2}\bigg] ,
\end{equation}
and hence the final state is squeezed in both degrees of freedom,
\begin{equation}
    \ket{\psi_f}=\exp\bigg\{GT_{\text{sq}}\bigg[\frac{(\hat{a}^\dagger)^2-\hat{a}^2}{2}\bigg]\bigg\}\ket{0}_a\otimes \exp\bigg\{GT_{\text{sq}}\bigg[\frac{(\hat{b}^\dagger)^2-\hat{b}^2}{2}\bigg]\bigg\}\ket{0}_b.
\end{equation}
In particular,
\begin{align}
    \begin{split}
        \exp\bigg\{-GT_{\text{sq}}\bigg[\frac{(\hat{b}^\dagger)^2-\hat{b}^2}{2}\bigg]\bigg\}(\hat{b}-\hat{b}^\dagger)\exp\bigg\{GT_{\text{sq}}\bigg[\frac{(\hat{b}^\dagger)^2-\hat{b}^2}{2}\bigg]\bigg\}=(\hat{b}-\hat{b}^\dagger)e^{-GT_{\text{sq}}}.
    \end{split}
\end{align}
To leading order in $N$, $\hat{S}_y=(\hat{S}^++\hat{S}^-)/2\approx i\sqrt{N}(\hat{b}-\hat{b}^\dagger)/2$, so
\begin{equation}
    \braket{\psi_f|\hat{S}_y^2|\psi_f}\approx - \frac{N}{4}\braket{\psi_f|(\hat{b}-\hat{b}^\dagger)^2|\psi_f}= - \frac{N}{4}\braket{\psi_0|(\hat{b}-\hat{b}^\dagger)^2|\psi_0}e^{-2GT_{\text{sq}}}=\frac{N}{4}e^{-2GT_{\text{sq}}}.
\end{equation}
Noting that $\braket{\hat{S}_x}\approx N/2$ to this same order of approximation, the amount of squeezing is
\begin{equation}\label{eqn:SuppSqueezingIdeal}
    \xi^2=\frac{N\braket{\hat{S}_y^2}}{\braket{\hat{S}_x}^2}=e^{-2GT_{\text{sq}}} .
\end{equation}
\subsection{Finite size effects}
To model finite size effects analytically, we add a phenomenological correction to Eq.~(\ref{eqn:SuppSqueezingIdeal}):
\begin{equation}\label{eqn:SuppPhenomenologicalCorrection}
    \xi^2_{\text{finite size}}=\xi^2+\frac{e^{2GT_{\text{sq}}}}{3N}=e^{-2GT_{\text{sq}}}+\frac{e^{2GT_{\text{sq}}}}{3N}
\end{equation}
The justification for the extra term is the following: to first approximation, curvature feeds the antisqueezed quadrature into the squeezed one, hence the factor $e^{2GT_{\text{sq}}}$, but is smaller by a factor of $1/N$ because the Holstein-Primakoff approximation is the leading order in an expansion in $1/N$. The prefactor is chosen so that the scaling agrees with the numerics. Minimizing the previous equation with respect to $T_{\text{sq}}$ leads to $\xi^2_{\text{opt}}=\sqrt{4/(3N)}$ and $GT_{\text{sq}}^{\text{opt}}=\log(3N)/4$, which correctly accounts for the $N$ scaling found numerically.

\section{Squeezing in the presence of decoherence}
As discussed in the main text, possible decoherence processes are dephasing with jump operator $\sqrt{\gamma_z}\hat{\sigma}_z^i/2$, spin flips, with jump operators $\sqrt{\gamma_+}\hat{\sigma}^+_i$ and $\sqrt{\gamma_-}\hat{\sigma}^-_i$, and boson decay, with jump operator $\sqrt{\kappa}\hat{a}$. In the case of the ion system we omit boson decay and also set $\gamma_+=\gamma_-$, which leads to the following evolution equation:
\begin{align}\begin{split}\label{eqn:SuppDecoherence0}
    \partial_t\hat{\rho}=-i\bigg[\delta\hat{a}^\dagger\hat{a}+\Omega\hat{S}_x+\frac{2g_{\text{ion}}}{\sqrt{N}}\hat{S}_z(\hat{a}^\dagger+\hat{a}),\hat{\rho}\bigg]+\frac{\gamma_z}{4}\sum_{i=1}^N\big(\hat\sigma_z^i\hat{\rho}\hat{\sigma}_z^i-\hat{\rho}\big)&+\frac{\gamma_+}{2}\sum_{i=1}^N\big(\hat\sigma_x^i\hat{\rho}\hat{\sigma}_x^i+\hat{\sigma}_y^i\hat{\rho}\hat{\sigma}_y^i-2\hat{\rho}\big),
\end{split}\end{align}
Conversely, the evolution equation for the cavity system is 
\begin{equation}\label{eqn:SuppCavityLiouvillian2}
    \partial_t\hat{\rho}=-i\Big[g_{\text{cav}}(\hat{a}\hat{S}^++\hat{a}^\dagger\hat{S}^-),\hat{\rho}\Big]+\kappa\bigg(\hat{a}\hat{\rho}\hat{a}^\dagger-\frac{1}{2}\{\hat{a}^\dagger\hat{a},\hat{\rho}\}\bigg)+\gamma\sum_{i=1}^N\big(\hat{\sigma}_i^-\hat{\rho}\hat{\sigma}_i^+-\frac{1}{2}\{\hat{\sigma}_i^+\hat{\sigma}_i^-,\hat{\rho}\}\big),
\end{equation}
% \begin{equation}\label{eqn:SuppCavityLiouvillian}
%     \partial_t\hat{\rho}=-i\Big[g_{\text{cav}}(\hat{a}\hat{S}^++\hat{a}^\dagger\hat{S}^-)+\omega_{\text{Rabi}}(t)\hat{S}_y,\hat{\rho}\Big]+\kappa\bigg(\hat{a}\hat{\rho}\hat{a}^\dagger-\frac{1}{2}\{\hat{a}^\dagger\hat{a},\hat{\rho}\}\bigg)+\gamma\sum_{i=1}^N\big(\hat{\sigma}_i^-\hat{\rho}\hat{\sigma}_i^+-\frac{1}{2}\{\hat{\sigma}_i^+\hat{\sigma}_i^-,\hat{\rho}\}\big),
% \end{equation}
where the only single particle decoherence process we have kept is spontaneous emission.

In the ion system we will resort to numerical methods described in the following sections to analyze the effects of dephasing and spin flips at intermediate values of $N$. In the cavity system, decoherence can be treated analytically in the large $N$ limit, as we now show.

\subsection{Cavity system}
The squeezing and transfer steps of the squeeze and transfer protocol are well described by Eq.~(\ref{eqn:SuppCavityLiouvillian2}). However, a $\pi$ pulse must be applied to toggle between them, which in the ideal case is assumed to be instantaneous. To account for a finite pulse duration time we add an extra Hamiltonian term to Eq.~(\ref{eqn:SuppCavityLiouvillian2}) describing an external Rabi drive, $\omega_{\text{Rabi}}(t)\hat{S}_y$. The evolution of the cavity system is then given by
\begin{equation}\label{eqn:SuppCavityLiouvillian}
    \partial_t\hat{\rho}=-i\Big[g_{\text{cav}}(\hat{a}\hat{S}^++\hat{a}^\dagger\hat{S}^-)+\omega_{\text{Rabi}}(t)\hat{S}_y,\hat{\rho}\Big]+\kappa\bigg(\hat{a}\hat{\rho}\hat{a}^\dagger-\frac{1}{2}\{\hat{a}^\dagger\hat{a},\hat{\rho}\}\bigg)+\gamma\sum_{i=1}^N\big(\hat{\sigma}_i^-\hat{\rho}\hat{\sigma}_i^+-\frac{1}{2}\{\hat{\sigma}_i^+\hat{\sigma}_i^-,\hat{\rho}\}\big).
\end{equation}
We allow for a time-dependent Rabi drive strength $\omega_{\text{Rabi}}(t)$ to account for the turning on/off of the pulse. For simplicity we first analyze the effects of a finite pulse duration in the presence of $\kappa$ without including $\gamma$, and then investigate the combined effects of finite $\gamma$ and $\kappa$ for an ideal $\pi$ pulse.

\subsubsection{Finite $\pi$ pulse duration}
Here we will look at the modifications introduced by a finite $\pi$ pulse duration in the presence of boson decay within the large $N$ approximation (we are setting $\gamma\to 0$ in this subsection). The full protocol involves a rotation halfway through, induced by $\omega_{\text{Rabi}}(t)$. In the limit $N\to\infty$, this rotation is mean-field like and is accompanied by the creation of a mean-field intracavity field. The mean field equations for this system, under the specific initial conditions of the squeeze and transfer protocol, are
\begin{equation}
    \dot{\theta}=\omega_{\text{Rabi}}(t)+2g\sqrt{N}\alpha(t)\hspace{2cm} \dot{\alpha}(t)=-\frac{\kappa\alpha(t)}{2}+\frac{g\sqrt{N}}{2}\sin\theta(t),
\end{equation}
where $\alpha=\braket{a}$ within the mean field approximation, and $\theta$ is the opening angle from the $z$ axis towards the mean field Bloch vector in the $xz$ plane (rotations generated by $\hat{S}_y$). Ultimately, we interested in are the fluctuations on top of the mean-field Bloch vector, not its mean field evolution. To study the former, we define the state in a moving frame, $\hat{\eta}$, through
\begin{equation}
    \hat{\rho}=e^{-i\theta(t)\hat{S}_y}e^{-i\alpha(t)\sqrt{N}(\hat{a}+\hat{a}^\dagger)}\hat{\eta}e^{i\theta(t)\hat{S}_y}e^{i\alpha(t)\sqrt{N}(\hat{a}+\hat{a}^\dagger)}.
\end{equation}
The unitaries extract the mean-field evolution so that in this moving frame the mean field Bloch vector is always pointing along $+z$ and the mean field cavity occupation is 0. All that remains are the dynamics of the quantum fluctuations. Since the state is always pointing along $+z$ in this frame, we can apply the Holstein-Primakoff approximation
\begin{equation}
\hat{S_z}\approx \frac{N}{2}-\hat{b}^\dagger\hat{b}\hspace{2cm}\hat{S}^+=\sqrt{N}\hat{b},
\end{equation}
The generator of time evolution becomes quadratic in boson operators, to lowest order in $1/N$:
\begin{equation}
    \partial_t\hat{\eta}=-ig_{\text{cav}}\sqrt{N}\Big[\cos(\theta/2)^2(\hat{a}\hat{b}+\hat{a}^\dagger\hat{b}^\dagger)+\sin(\theta/2)^2(\hat{a}^\dagger\hat{b}+\hat{b}^\dagger\hat{a}),\hat{\eta}\Big]+\kappa\bigg(\hat{a}\hat{\eta}\hat{a}^\dagger-\frac{1}{2}\{\hat{a}^\dagger\hat{a},\hat{\eta}\}\bigg),
\end{equation}
and we have omitted the time dependence of $\theta(t)$ for notational clarity. This evolution equation induces equations of motion for expectation values that close among themselves. We are interested in quadratic fluctuations, so we define the matrix
\begin{equation}
    M(t)=\begin{pmatrix}
    \braket{\hat{a}\hat{a}}&\braket{\hat{a}\hat{b}}&\braket{\hat{a}\hat{a}^\dagger}&\braket{\hat{a}\hat{b}^\dagger}\\
    \braket{\hat{b}\hat{a}}&\braket{\hat{b}\hat{b}}&\braket{\hat{b}\hat{a}^\dagger}&\braket{\hat{b}\hat{b}^\dagger}\\
    \braket{\hat{a}^\dagger\hat{a}}&\braket{\hat{a}^\dagger\hat{b}}&\braket{\hat{a}^\dagger\hat{a}^\dagger}&\braket{\hat{a}^\dagger\hat{b}^\dagger}\\
    \braket{\hat{b}^\dagger\hat{a}}&\braket{\hat{b}^\dagger\hat{b}}&\braket{\hat{b}^\dagger\hat{a}^\dagger}&\braket{\hat{b}^\dagger\hat{b}^\dagger}.
    \end{pmatrix}
\end{equation}
It satisfies the matrix equation of motion
\begin{equation}\label{eqn:SuppMatrixEqn}
    \partial_t{M}(t)=FM(t)+M(t)F^T+D,
\end{equation}
where 
\begin{equation}
    F=-ig_{\text{cav}}\sqrt{N}\begin{pmatrix}
    -i\kappa/(2g_{\text{cav}}\sqrt{N})&\sin(\theta/2)^2&0&\cos(\theta/2)^2\\[5pt]
    \sin(\theta/2)^2&0&\cos(\theta/2)^2&0\\[5pt]
    0&-\cos(\theta/2)^2&-i\kappa/(2g_{\text{cav}}\sqrt{N})&-\sin(\theta/2)^2\\[5pt]
    -\cos(\theta/2)^2&0&-\sin(\theta/2)^2&0
    \end{pmatrix}
\end{equation}
is a time-dependent matrix [through $\theta(t)$] and
\begin{equation}
    D=\begin{pmatrix}0&0&\kappa&0\\
    0&0&0&0\\
    0&0&0&0\\
    0&0&0&0
    \end{pmatrix}\hspace{2cm}M(0)=\begin{pmatrix}0&0&1&0\\
    0&0&0&1\\
    0&0&0&0\\
    0&0&0&0
    \end{pmatrix}.
\end{equation}
Spin squeezing as a function of time can then be obtained from $M_{24}=\braket{\hat{b}\hat{b}^\dagger}$, $M_{42}=\braket{\hat{b}^\dagger\hat{b}}$ and $M_{22}=\braket{\hat{b}\hat{b}}$ as
\begin{equation}\label{eqn:SuppSpinSqueezing}
    \xi^2=\text{Min}_{\nu}\braket{(\hat{b}e^{i\nu}+\hat{b}^\dagger e^{-i\nu})^2}=M_{24}+M_{42}-2\text{Abs}(M_{22}),
\end{equation}
while the optimal $\nu$ determines the quadrature along which the state is spin squeezed. Given the structure of $F(t)$, $\nu=\pi/2$ or $\nu=0$ in general, so the squeezed variable is always $\hat{S}_y$ of $\hat{S}_x$ in the rotated frame.

% \begin{equation}\label{eqn:Supp_cavity}
%     \xi^2=\frac{\kappa^2}{16g^2}e^{2g_{\text{cav}}\sqrt{N}T_{\text{sq}}}e^{-\kappa T_{\text{sq}}/2}+e^{-2g_{\text{cav}}\sqrt{N}T_{\text{sq}}}e^{-\kappa T_{\text{sq}}/2}+\frac{\kappa}{2g_{\text{cav}}\sqrt{N}}(e^{-\kappa T_{\text{sq}}/2}-1)+\frac{\pi\kappa}{8g_{\text{cav}}\sqrt{N}}.
% \end{equation}
% Optimizing this expression with respect to $T_{\text{sq}}$ and to lowest order in $\kappa/g$ leads to $2g_{\text{cav}}\sqrt{N}T_{\text{sq}}^{\text{opt}}=\log(4g_{\text{cav}}\sqrt{N}/\kappa)$ and thus an optimal sensitivity of 
% \begin{equation}
%     \xi^2_{\text{opt}}\approx\frac{\kappa}{g_{\text{cav}}\sqrt{N}}\bigg[\frac{\pi}{8}-\frac{1}{2}+e^{-\frac{\kappa}{4g_{\text{cav}}\sqrt{N}}\log(4g_{\text{cav}}\sqrt{N}/\kappa)}\bigg]\approx 0.89\frac{\kappa}{g_{\text{cav}}\sqrt{N}}-\frac{\kappa^2}{4g_{\text{cav}}^2N}\log(4g_{\text{cav}}\sqrt{N}/\kappa).
% \end{equation}
% The resulting scaling with $\kappa$ is very favorable for small $\kappa/(g_{\text{cav}}\sqrt{N})$ and comes about because of the exponentially fast squeezing of the coherent dynamics. Notice also that we are not optimizing with respect to $T_{\text{tr}}$ and instead using the value calculated from the purely coherent dynamics. It is possible to gain an extra $3$dB of squeezing if we optimize over $T_{\text{tr}}$ too.
We assume the Rabi drive is turned on for a finite time $T_{\text{drive}}$.  The sequence of steps is as follows: (1) $\omega_{\text{Rabi}}=0$ for a time $T_{\text{sq}}$, where the interactions create the two-mode squeezed state. (2) A finite Rabi drive, with strength $\omega_{\text{Rabi}}$, is turned on for a time $T_{\text{drive}}$ and then switched off. (3) The state is left to evolve under atom-cavity interactions for a time $T_{\text{tr}}$. 

\begin{figure}
    \centering
    \includegraphics[width=\textwidth]{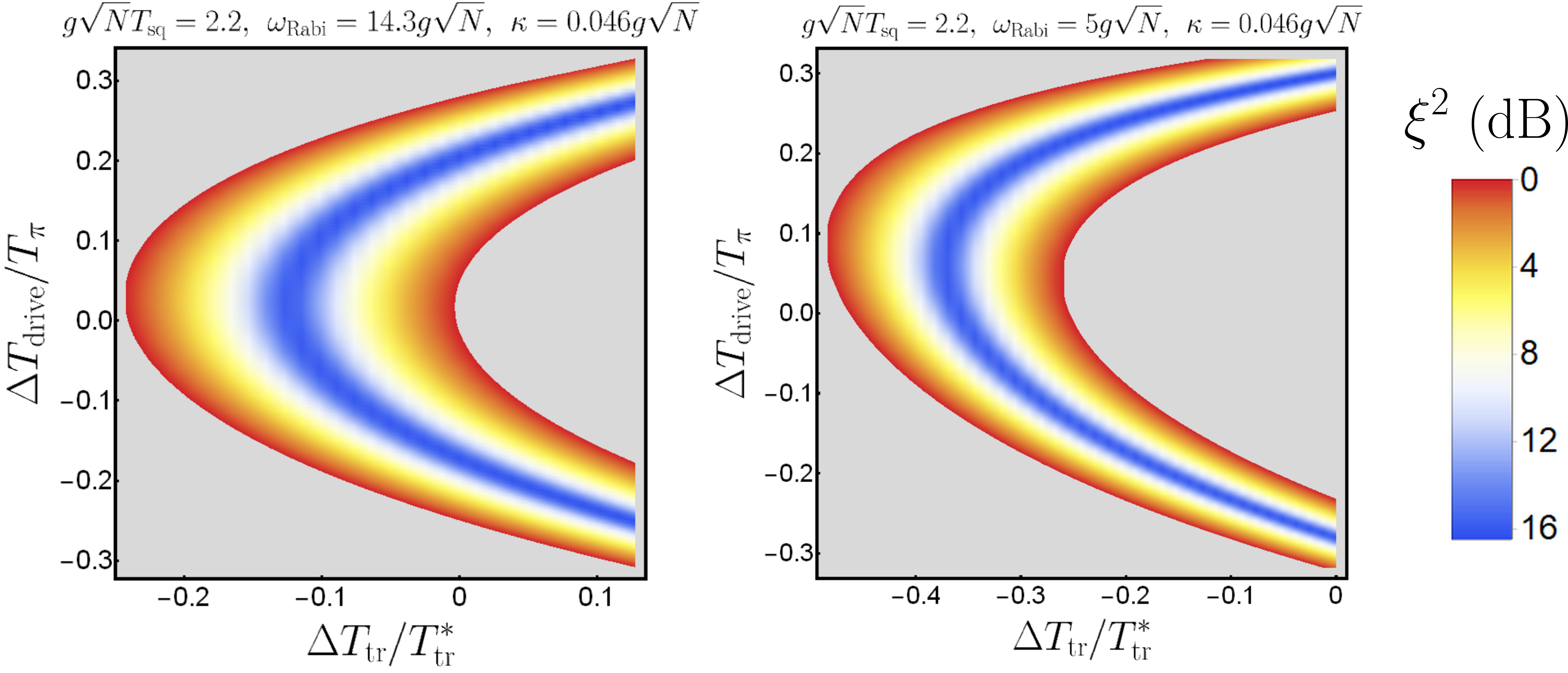}
    \caption{Color map of squeezing as a function of $T_{\text{tr}}$ and $T_{\text{drive}}$ for fixed $g_{\text{cav}}\sqrt{N}T_{\text{sq}}=2.2$ and $\omega_{\text{Rabi}}\approx 15 g_{\text{cav}}\sqrt{N}$ (left) and $\omega_{\text{Rabi}}= 5 g_{\text{cav}}\sqrt{N}$ (right). The horizontal axes are the deviation $\Delta T_{\text{tr}}=T_{\text{tr}}-T_{\text{tr}}^*$ from the ideal result $g\sqrt{N}T_{\text{tr}}^*=\pi/4$ in units of the ideal transfer time. The vertical axis is the deviation $\Delta T_{\text{drive}}=T_{\text{drive}}-T_{\pi}$ from the ideal duration of the $\pi$ pulse, $T_{\pi}=\pi/\omega_{\text{Rabi}}$, in units of $T_{\pi}$. Gray regions on the outskirts are oversqueezed.}
    \label{fig:SuppFinitePulseTime}
\end{figure}

In the ideal scenario, $\omega_\text{Rabi}=\infty$, $T_{\text{drive}}=0$ and $g_{\text{cav}}\sqrt{N}T_{\text{tr}}^*=\pi/4$. However, in the presence of a finite Rabi drive strength, for given $T_{\text{sq}}$ we can optimize over $T_{\text{drive}}$ and $T_{\text{tr}}$ to obtain the best possible squeezing and assess whether a finite pulse duration limits the amount of achievable squeezing. To investigate this, we do numerical simulations of Eq.~(\ref{eqn:SuppMatrixEqn}) for fixed $g_{\text{cav}}\sqrt{N}T_{\text{sq}}=2.2$, and two values of $\omega_{\text{Rabi}}$: $\omega_{\text{Rabi}}\approx 15 g_{\text{cav}}\sqrt{N}$ and $\omega_{\text{Rabi}}= 5 g_{\text{cav}}\sqrt{N}$, as we vary $T_{\text{tr}}$ and $T_{\text{drive}}$. We show the results in Fig.~\ref{fig:SuppFinitePulseTime}, which indicate that no squeezing is lost.

\subsubsection{Inclusion of spontaneous emission}\label{sec:SuppSpontaneousEmission}
When spontaneous emission is included, the Holstein-Primakoff approximation is no longer valid because the state is no longer restricted to the Dicke manifold. Nevertheless, the large $N$ approximation can still be implemented at the level of the equations of motion. Similar to the previous section, the idea is to first compute the mean field evolution of the system and then analyze the equations for fluctuations with respect to the mean field background. To simplify things we will thus work assuming that the $\pi$ pulse is instantaneous. The mean field equations are then
\begin{align}\begin{split}
    \dot{\alpha}_{\text{mf}}&=-ig_{\text{cav}}s_{\text{mf}}-\kappa\alpha_{\text{mf}}/2\\
    \dot{s}_{\text{mf}}&=2ig_{\text{cav}}z_{\text{mf}}\alpha_{\text{mf}}-\gamma s_{\text{mf}}/2\\
    \dot{z}_{\text{mf}}&=-ig_{\text{cav}}(s_{\text{mf}}^*\alpha_{\text{mf}}-s_{\text{mf}}\alpha_{\text{mf}}^*)-\gamma \bigg(z_{\text{mf}}+\frac{N}{2}\bigg),
\end{split}\end{align}
obtained by factorizing expectation values in the Heisenberg equations of motion for $\braket{\hat{a}},\braket{\hat{S}^-},\braket{\hat{S}_z}$. For spins initially pointing along $+z$ and no boson occupation, these equations have solution $\alpha_{\text{mf}}=0,s_{\text{mf}}=0$ (thus $\hat{S}^-$ and $\hat{a}$ describe fluctuations), while
\begin{equation}
    z_{\text{mf}}=\begin{dcases}
        Ne^{-\gamma t}-N/2 & 0\leq t<T_{\text{sq}} \\
        N(1-e^{-\gamma T_{\text{sq}}})e^{-\gamma(t-T_{\text{sq}})} -N/2& T_{\text{sq}}<t\leq T_{\text{sq}}+T_{\text{tr}} \\
    \end{dcases}
\end{equation}
evolves in two stages. For $0\leq t<T_{\text{sq}}$, $z_{\text{mf}}$ starts at $N/2$ and is reduced by spontaneous emission to the value $Ne^{-\gamma T_{\text{sq}}}-N/2$. The instantaneous $\pi$ pulse transforms this into $-Ne^{-\gamma T_{\text{sq}}}+N/2$, which is the starting value of $z_{\text{mf}}$ in the time window $T_{\text{sq}}\leq t<T_{\text{sq}}+T_{\text{tr}}$. 

The equations of motion for the relevant fluctuations can be written efficiently by introducing the abstract vectors
\begin{equation}
    \hat{o}=\begin{pmatrix}
        \frac{\hat{S}^-}{\sqrt{N}}\\[5pt]
        \hat{a}
        \end{pmatrix}
        \hspace{2cm}
        \hat{o}^\dagger=\begin{pmatrix}
        \frac{\hat{S}^+}{\sqrt{N}}\\[5pt]
        \hat{a}^\dagger
        \end{pmatrix}
\end{equation}
and considering tensor products in this abstract vector space
\begin{equation}
   \hat{o}\otimes\hat{o}=\begin{pmatrix}
        \frac{\hat{S}^-}{\sqrt{N}}\\[5pt]
        \hat{a}
        \end{pmatrix}\otimes \begin{pmatrix}
        \frac{\hat{S}^-}{\sqrt{N}}\\[5pt]
        \hat{a}
        \end{pmatrix}\hspace{2cm}\hat{o}^\dagger\otimes \hat{o}=\begin{pmatrix}
        \frac{\hat{S}^+}{\sqrt{N}}\\[5pt]
        \hat{a}^\dagger
        \end{pmatrix}\otimes \begin{pmatrix}
        \frac{\hat{S}^-}{\sqrt{N}}\\[5pt]
        \hat{a}
        \end{pmatrix}.
\end{equation}
For clarity, we remark that if $A$ and $B$ are $2\times 2$ matrices, then $(A\otimes B)(\hat{o}\otimes\hat{o})=(A\hat{o})\otimes(B\hat{o})$. In terms of this construction, the equations of motion take the form
\begin{align}\label{eqn:SuppFluctuations}
    \begin{split}
        \partial_t \braket{\hat{o}\otimes\hat{o}}&=[G(t)\otimes \mathds{1}+\mathds{1}\otimes G(t)]\braket{\hat{o}\otimes\hat{o}}
    \\[10pt]
    &-2ig_{\text{cav}}\sqrt{N}\frac{\braket{\hat{a}\hat{S}^-\delta\hat{S}_z}+\braket{\hat{a}\delta\hat{S}_z\hat{S}_-}}{N^{3/2}}\begin{pmatrix}
        1\\[5pt]
        0
        \end{pmatrix}\otimes \begin{pmatrix}
        1\\[5pt]
        0
        \end{pmatrix}+2ig_{\text{cav}}\sqrt{N}\frac{\braket{\hat{a}^2\delta\hat{S}_z}}{N^{3/2}}\Bigg[\begin{pmatrix}
        1\\[5pt]
        0
        \end{pmatrix}\otimes \begin{pmatrix}
        0\\[5pt]
        1
        \end{pmatrix}+\begin{pmatrix}
        0\\[5pt]
        1
        \end{pmatrix}\otimes \begin{pmatrix}
        1\\[5pt]
        0
        \end{pmatrix}\Bigg],
\end{split}\end{align}
where $\delta\hat{S}_z\equiv \hat{S}_z-z_{\text{mf}}$ and
\begin{equation}
    G(t)=\begin{pmatrix}
    -\gamma/2&ig_{\text{cav}}\sqrt{N}\big[z_{\text{mf}}/(N/2)\big]\\[5pt]
    -ig_{\text{cav}}\sqrt{N}&-\kappa/2
    \end{pmatrix}
\end{equation}
is a time-dependent (through $z_{\text{mf}}$) matrix. For the kind of states we are considering, $\hat{S}^-\sim\sqrt{N}$, $\hat{a}\sim 1$, $\delta\hat{S}_z \sim \sqrt{N}$, so the second and third lines are a factor $\sqrt{N}$ smaller than the first one, and can thus be neglected in the large $N$ limit. Under this approximation, it follows that
\begin{align}
    \begin{split}
      \braket{\hat{o}\otimes\hat{o}}_t=P(t,t_0)\otimes P(t,t_0)\braket{\hat{o}\otimes\hat{o}}_{t_0}
    \end{split}
\end{align}
where $P(t,t_0)$ is a $2\times2$ matrix propagator, defined by the equations $\partial_t{P}=G(t)P$, $P(t_0,t_0)=\mathds{1}$. Similarly,
\begin{equation}\label{eqn:SuppPropagator2}
    \braket{\hat{o}^\dagger\otimes\hat{o}}_t=P(t,t_0)^*\otimes P(t,t_0)\braket{\hat{o}^\dagger\otimes\hat{o}},
\end{equation}
where, crucially, $\hat{S}^+$ and $\hat{a}^\dagger$ are to the left of $\hat{S}^-$ and $\hat{a}$. This last equation does not hold true in the reverse order. Of particular importance will be $L=P(T_{\text{sq}},0)$, the propagator from time $t=0$ to the squeezing time $T_{\text{sq}}$, immediately before the $\pi$ pulse; and $M=P(T_{\text{sq}}+T_{\text{tr}},T_{\text{sq}})$, the propagator from time $t=T_{\text{sq}}$, immediately after the $\pi$ pulse, to the end of the protocol. To compute spin squeezing, we need to calculate the expectation values $\braket{(\hat{S}^-)^2}$ and $\braket{\hat{S}^+\hat{S}^-}$ at the end of the protocol, which involves various steps. Let us begin with $\braket{(\hat{S}^-)^2}$:
\begin{align}
    \begin{split}
        \bigg\langle\frac{(\hat{S}^-)^2}{N}\bigg\rangle_{\text{End}}&=\Bigg\langle \Bigg[v_1M\begin{pmatrix} \frac{\hat{S}^-}{\sqrt{N}}\\[5pt]
        \hat{a}
        \end{pmatrix}\Bigg]\Bigg[v_1M\begin{pmatrix} \frac{\hat{S}^-}{\sqrt{N}}\\[5pt]
        \hat{a}
        \end{pmatrix}\Bigg]\Bigg\rangle_{\text{After $\pi$}}=\Bigg\langle \Bigg[v_1M\begin{pmatrix} \frac{-\hat{S}^+}{\sqrt{N}}\\[5pt]
        \hat{a}
        \end{pmatrix}\Bigg]\Bigg[v_1M\begin{pmatrix} \frac{-\hat{S}^+}{\sqrt{N}}\\[5pt]
        \hat{a}
        \end{pmatrix}\Bigg]\Bigg\rangle_{\text{Before $\pi$}}
    \end{split}
\end{align}
In the first equality we used $M$ to express the final expectation value in terms of the expectation value immediately after the $\pi$ pulse and $v_1=(1,0)$, $v_2=(0,1)$ are vectors used to project into specific components of $\hat{o}$ and $\hat{o}^\dagger$. In the second equality, we used the fact that a $\pi$ pulse along $y$ maps $\hat{S}^{\pm}\to-\hat{S}^{\mp}$. We now need to relate the expectation values immediately before the $\pi$ pulse to those at the beginning of the protocol using $L$. First, we note that $L$ will mix $\braket{(\hat{S}^+)^2}$ with $\braket{\hat{S}^+\hat{a}^\dagger}$ and $\braket{(\hat{a}^\dagger)^2}$, both of which are $0$ initially. The same happens with $\braket{(\hat{a}^\dagger)^2}$. Thus we only need
\begin{equation}
    \bigg\langle\frac{\hat{S}^+\hat{a}}{\sqrt{N}}\bigg\rangle_{\text{Before $\pi$}}=\Bigg\langle \Bigg[v_1 L^*\begin{pmatrix} \frac{\hat{S}^+}{\sqrt{N}}\\[5pt]
        \hat{a}^\dagger
        \end{pmatrix}\Bigg]\Bigg[v_2L\begin{pmatrix} \frac{\hat{S}^-}{\sqrt{N}}\\[5pt]
        \hat{a}
        \end{pmatrix}\Bigg]\Bigg\rangle_{\text{initial}}=L_{11}^* L_{21}.
\end{equation}
The final result is expressed in terms of matrix elements of $M$ and $L$
\begin{equation}
   \boxed{ \bigg\langle\frac{(\hat{S}^-)^2}{N}\bigg\rangle_{\text{End}}=-2L_{11}^*L_{21}M_{12}M_{11}}
\end{equation}
The expression for $\braket{\hat{S}^+\hat{S}^-}$ is obtained in a similar manner, except for one extra complication, which we point out explicitly:
\begin{align}
    \begin{split}
        \bigg\langle\frac{\hat{S}^+\hat{S}^-}{N}\bigg\rangle_{\text{End}}&=\Bigg\langle \Bigg[v_1M^*\begin{pmatrix} \frac{\hat{S}^+}{\sqrt{N}}\\[5pt]
        \hat{a}^\dagger
        \end{pmatrix}\Bigg]\Bigg[v_1M\begin{pmatrix} \frac{\hat{S}^-}{\sqrt{N}}\\[5pt]
        \hat{a}
        \end{pmatrix}\Bigg]\Bigg\rangle_{\text{After $\pi$}}=\Bigg\langle \Bigg[v_1M^*\begin{pmatrix} \frac{-\hat{S}^-}{\sqrt{N}}\\[5pt]
        \hat{a}^\dagger
        \end{pmatrix}\Bigg]\Bigg[v_1M\begin{pmatrix} \frac{-\hat{S}^+}{\sqrt{N}}\\[5pt]
        \hat{a}
        \end{pmatrix}\Bigg]\Bigg\rangle_{\text{Before $\pi$}}\\[10pt]
        &=|M_{11}|^2\bigg\langle\frac{\hat{S}^-\hat{S}^+}{N}\bigg\rangle_{\text{Before $\pi$}}+|M_{12}|^2\braket{\hat{a}^\dagger\hat{a}}_{\text{Before $\pi$}}\\[10pt]
        &=|M_{11}|^2\bigg\langle\frac{\hat{S}^+\hat{S}^-}{N}\bigg\rangle_{\text{Before $\pi$}}+|M_{12}|^2\braket{\hat{a}^\dagger\hat{a}}_{\text{Before $\pi$}}-2|M_{11}|^2\bigg\langle\frac{\hat{S}_z}{N}\bigg\rangle_{\text{Before $\pi$}}.
    \end{split}
\end{align}
The second line is obtained after replacing $\braket{\hat{a}^\dagger\hat{S}^+}_{\text{Before $\pi$}}=0$ and $\braket{\hat{a}\hat{S}^-}_{\text{Before $\pi$}}=0$. To apply Eq.~(\ref{eqn:SuppPropagator2}) the operators have to be in the order $\hat{S}^+\hat{S}^-$, so in the third line we used equal time commutation relations, which contribute an extra term. Applying similar manipulations as before, we obtain
\begin{equation}
    \boxed{\bigg\langle\frac{\hat{S}^+\hat{S}^-}{N}\bigg\rangle_{\text{End}}=|M_{11}|^2|L_{11}|^2+|M_{12}|^2|L_{21}|^2-(2e^{-\gamma T_{\text{sq}}}-1)|M_{11}|^2}.
\end{equation}
Spin squeezing is obtained applying arguments identical to Eq.~(\ref{eqn:SuppSpinSqueezing}). Thus, the minimal variance and the contrast at the end of the protocol are given by
\begin{align}\begin{split}
    \frac{\text{Var}_{\text{min}}}{N/4}&=\frac{1}{N}\bigg[\braket{\hat{S}^+\hat{S}^-}_{\text{End}}+\braket{\hat{S}^-\hat{S}^+}_{\text{End}}-2\Big|\braket{\hat{S}^-\hat{S}^-}_{\text{End}}\Big|\bigg]\\
    \frac{\text{Var}_{\text{min}}}{N/4}&=-2\big(2e^{-\gamma T_{\text{sq}}}-1\big)|M_{11}|^2+\Big[1-2\big(1-e^{-\gamma T_{\text{sq}}}\big)e^{-\gamma T_{\text{tr}}}\Big]+2\Big(|M_{11}||L_{11}|-|M_{12}||L_{21}|\Big)^2\\
   \text{Contrast}&=1-2(1-e^{-\gamma T_{\text{sq}}})e^{-\gamma T_{\text{tr}}}.
\end{split}\end{align}
Since $G(t)$ is time dependent there is no closed form expression for $L$ and $M$, so we have to resort to numerical simulations, though only of a $2\times 2$ matrix equation.

We can do analytics when $\gamma \to 0$, in which case $G(t)$ becomes time-independent. For small $\kappa\ll g_{\text{cav}}\sqrt{N}$, we expect the optimal transfer time to remain close to $\pi/4$ so that the quantity $\epsilon =g\sqrt{N}T_{\text{tr}}-\pi/4$ is small. Calculating $M$ and $L$ to lowest order in $\kappa/(g_{\text{cav}}\sqrt{N})$ and $\epsilon$ leads to
\begin{equation}\label{eqn:Supp_cavity}
    \xi^2=\frac{\pi \kappa}{8g\sqrt{N}}+1-\bigg(1+\frac{\kappa}{4g\sqrt{N}}-\epsilon\bigg)^2+\bigg[e^{g\sqrt{N}T_{\text{sq}}}\bigg(\frac{\kappa}{4g\sqrt{N}}-\epsilon\bigg)+e^{-g\sqrt{N}T_{\text{sq}}}\bigg]^2.
\end{equation}
The minimum squeezing is obtained asymptotically at $\epsilon=\kappa/4(g\sqrt{N})$ and $T_{\text{sq}}\to\infty$ (the optimal $T_{\text{sq}}$ is determined by higher order terms in $\kappa$ that we have neglected but do not change the achievable squeezing significantly), with optimal value is $\xi^2_{\text{opt}}\approx 0.4 \kappa/(g\sqrt{N})$. In practice, working with $2g\sqrt{N}T_{\text{sq}}\gtrsim \log\big(2.5g\sqrt{N}/\kappa\big)$ is enough.

% \subsubsection{Finite size effects}
% To model finite size effects analytically, we add the phenomenological correction discussed in Eq.~(\ref{eqn:SuppPhenomenologicalCorrection}) to Eq.~(\ref{eqn:Supp_cavity}):
% \begin{equation}
%     \xi^2_{\text{finite size}}=\xi^2+\frac{e^{2g_{\text{cav}}\sqrt{N}T_{\text{sq}}}}{3N}.
% \end{equation}
% We numerically minimize $\xi^2_{\text{finite size}}$ using $\xi^2$ from Eq.~(\ref{eqn:Supp_cavity}) to obtain Fig.~3 in the main text.
% In the absence of boson decay ($\kappa=0$), optimizing $\xi^2_{\text{finite size}}=e^{-2GT_{\text{sq}}}+e^{2GT_{\text{sq}}}/(3N)$ leads to $\xi^2_{\text{opt}}=\sqrt{4/(3N)}$ and $GT_{\text{sq}}^{\text{opt}}=\log(3N)/4$, which correctly accounts for the $N$ scaling found numerically. We thus expect it to give a qualitatively correct description even when $\kappa\neq 0$, so 
% We then numerically minimize this analytical expression over $g T_{\text{sq}}$ at fixed $g T_{\text{tr}}=\pi/4$ and $\kappa/g$. The resulting minimum sensitivity as a function of $\kappa/g$ is plotted in Fig.~4 in the main text.

\section{Numerical simulations (exact)}\label{sec:ExactNumerics}
\Rev{To examine the effects of finite $N$ on our protocols, as well as the presence of a finite detuning $\delta$ in the ion implementation}, we efficiently solve the time-dependent Schr{\"o}dinger equation utilizing a truncated Fock space with maximum occupation $n_{\mathrm{max}}$ for the boson degree of freedom, in combination with a Krylov-subspace projection method \cite{Sidje_1998}. Owing to the need to optimize our protocol over a wide range of parameters, and simulate Eq.~(4) for several different dynamical regimes, we take the same approach as in \cite{Lewis-Swan2021}, implementing a dynamic cutoff $n_{\mathrm{max}}$ that grows/shrinks to optimize our simulation time, while also accommodating any growth in the boson population. This allows us to avoid extensive benchmarking to identify a suitable cutoff $n_{\textrm{max}}$ for each set of parameters, while also minimizing computational resources to allow efficient solutions for large system sizes.

\begin{figure}[tbh]\centering
\includegraphics[width=0.78\textwidth]{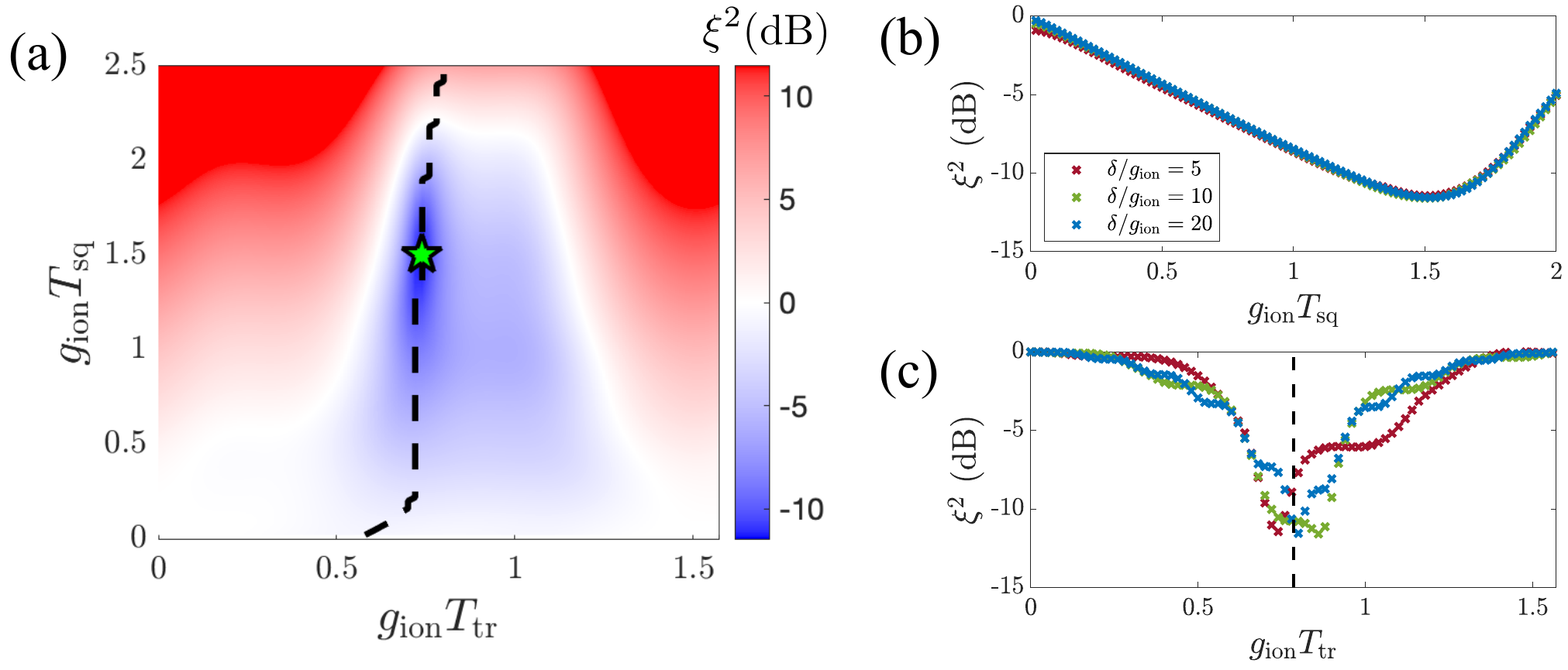}
\caption{Squeeze and transfer protocol in an ion setup for a range of protocol times with $\delta = 5g_{\mathrm{ion}}$ and $N=250$. a) We plot the spin squeezing parameter $\xi^2$, shown in decibels (dB), for a range of $T_{\mathrm{sq}}$ and $T_{\mathrm{tr}}$. The dashed line indicates the minimum squeezing for each fixed $T_{\mathrm{sq}}$ when optimized over $T_{\mathrm{tr}}$, which is shown in Fig.~2 of the main text. The green star indicates the optimal protocol with $g_{\mathrm{ion}}T_{\mathrm{sq}} \approx 1.50$ and $g_{\mathrm{ion}}T_{\mathrm{tr}} \approx 0.74$, yielding $11.4$ dB of squeezing. These are close to the predicted values of $g_{\mathrm{ion}}T_{\mathrm{sq}} = (\log N)/4 \approx 1.38$ and $g_{\mathrm{ion}}T_{\mathrm{tr}} = \pi/4 \approx 0.79$.
For the squeezing protocol with $N=250$, we plot the attainable spin squeezing b) as a function of $T_{\mathrm{sq}}$ when optimized over the BS evolution time $T_{\mathrm{tr}}$ and c) as a function of $T_{\mathrm{tr}}$ when optimized over the TMS evolution time $T_{\mathrm{sq}}$ for various values of $\delta = |\Omega|$.}
\label{fig:supp1}
\end{figure}

%For the squeezing protocol, we apply the TMS Hamiltonian to our initial state $\ket{\psi(0)}=\ket{N/2_x}\otimes\ket{0}$ for various squeezing times $T_{\text{sq}}$ to obtain $e^{-i\hat{H}_{\text{TMS}}T_{\text{sq}}}\ket{\psi(0)}$, and then apply the BS Hamiltonian for a time $T_{\text{tr}}$.
In Fig.~\ref{fig:supp1}, we show the results of our calculations for the squeezing and time-reversal protocols with $\delta = |\Omega| = 5g_{\mathrm{ion}}$, as used in the main text, and $N=250$, plotted over the full range of possible protocol times we consider. By optimizing these results over the calculated times (and similarly for other $N$), we obtain the results shown in Fig.~2 of the main text. We also plot the results of our squeezing protocol for different $\delta/g_{\mathrm{ion}}$, finding that the optimal values of $T_{\mathrm{sq}}$ and $T_{\mathrm{tr}}$, as well as the corresponding squeezing, remain robust to variations in this parameter. As $\delta$ increases, the optimal value of $g_{\mathrm{ion}}T_{\mathrm{tr}}$ approaches $\pi/4$.

For comparisons with the corresponding OAT protocol in Fig.~2 of the main text, we simulate Eq.~(4) of the main text with $\Omega = 0$ for various values of $\delta/g_{\mathrm{ion}}$, optimizing over this parameter for each value of $T$. While OAT dynamics in our spin-boson model are technically only realized at the stroboscopic times $T = 2\pi n/\delta$ for integer $n$, in between these times we always observe a degradation in the amount of achievable squeezing when compared to the corresponding OAT dynamics. As a result, for any given protocol time $T$, we find that optimizing the squeezing over $\delta$ always results in an optimal value of $\delta \approx 2\pi n/T$ that naturally satisfies our stroboscopic condition.

%..., restricting our analysis of the squeezing to times set by stroboscopic condition $T = 2\pi n/\delta$. In Fig.~2 of the main text, for each time $gT$, we plot the corresponding $\xi^2$ at that time obtained from the dynamics with $\delta = 2\pi/T$ (we note that this provides essentially the same results when optimizing over all $T$ and $\delta$ without regard to the stroboscopic condition).

%For the time-reversal protocol, we compute ... commutator matrix. two-time correlation matrix. optimization. for various $T_sq$, $T_tr$ and optimize. Plot unoptimized results for N=???, $\delta = 5g$. Indicate perfect time-reversal, optimized over-reversal. Plot of optimized $T_1 - T_2$, comparison for different $\delta/g$.

\section{Numerical simulations (semiclassical)}
To examine the effect of dissipation for the ion system, we make use the semiclassical dissipative discrete truncated Wigner approximation (DDTWA) \cite{Huber2022} to simulate the dynamics of the master equation in Eq.~\eqref{eqn:SuppDecoherence0}. Here, we briefly review the basic method introduced in \cite{Huber2021} and provide benchmarks of our protocol for small system sizes.% Within this framework, coherent dynamics of the system are computed by solving the associated classical equations of motion, in conjunction with random initial conditions sampled from a phase space distribution of the corresponding initial state. Incoherent terms in our master equation may be accounted for by the addition of stochastic noise terms to our classical equations of motion; these noise terms are assumed to be uncorrelated in time, in analogy to the Markovian approximation invoked to produce the master equation. Quantum expectation values are then obtained by averaging appropriate quantities over our collection of trajectories. Given the generic nonlinear nature of our classical equations of motion, this averaging produces results beyond mean-field theory that take into account the effect of the quantum noise distribution on our dynamics.

We begin by formulating a semiclassical description of our system, introducing classical variables $\alpha$ and $s_i^\mu$ corresponding to the values of $\langle \hat{a}\rangle$ and $\langle \hat{\sigma}_i^\mu\rangle/2$, where $\mu = x,y,z$ and $1 \leq i \leq N$. For the initial state $\ket{(N/2)_x}\otimes\ket{0}_a$, the bosonic Wigner function for $\alpha$ takes the form
\begin{align}
    W(\alpha) = \frac{2}{\pi}e^{-|\alpha|^2},
\end{align}
while for each of the spin-1/2 particles in our system, we form the discrete Wigner function
\begin{align}
    W(\vec{s}_i) = \frac{1}{4}\delta(s_i^x - 1/2)\Big[\delta(s_i^y - 1/2) + \delta(s_i^y + 1/2)\Big]\Big[\delta(s_i^z - 1/2) + \delta(s_i^z + 1/2)\Big].
\end{align}
For each spin, this amounts to the four phase space points $(s_i^x,s_i^y,s_i^z) = (0.5,\pm 0.5, \pm 0.5)$ each occurring with equal probability $1/4$.

Within this framework, coherent dynamics of the system are obtained by solving the associated classical equations of motion, in conjunction with randomly sampling initial values for $\alpha$, $(s_i^x,s_i^y,s_i^z)_{1\leq i\leq N}$ according to the above distributions. Incoherent terms in our master equation may be accounted for by the addition of stochastic noise terms to our classical equations of motion; these noise terms are assumed to be uncorrelated in time, in analogy to the Markovian approximation invoked to produce the master equation. While dissipation can in principle also be accounted for by a mean-field approach, the stochastic treatment introduced in \cite{Huber2021} has the added benefit that our total spin length for each spin-1/2 particle is conserved for any given classical trajectory, drastically improving the accuracy of our results. Applying this to Eq.~\eqref{eqn:SuppDecoherence0}, we have the resulting Stratonovich stochastic differential equations
%To simulate the dynamics of this initial state within DDTWA, we sample initial values for $\alpha$, $(s_i^x,s_i^y,s_i^z)_{1\leq i\leq N}$ according to the above distributions, and evolve these via the Stratonovich stochastic differential equations
\begin{align}
\begin{split}
    d\alpha &= \bigg[-i\delta\alpha - \frac{2ig_{\mathrm{ion}}}{\sqrt{N}}\sum_i s_i^z \bigg] dt\\
    ds_i^x &=  \frac{-2g_{\mathrm{ion}}}{\sqrt{N}}\left(\alpha + \alpha^*\right) s^y_i dt + \sqrt{2\gamma_+}s_i^z dW_i^y - \sqrt{\gamma_z}s_i^ydW_i^z\\
    ds_i^y &= \bigg[\frac{2g_{\mathrm{ion}}}{\sqrt{N}}\left(\alpha + \alpha^*\right) s^x_i -\Omega s^z_i \bigg] dt - \sqrt{2\gamma_+}s_i^z dW_i^x + \sqrt{\gamma_z}s_i^xdW_i^z\\
    ds_i^z &= \Omega s^y_i dt + \sqrt{2\gamma_+}\left(s_i^y dW_i^x - s_i^xdW_i^y\right)
    \end{split}
\end{align}
for independent Wiener increments $dW_i^{\mu} \equiv dW_i^\mu(t)$, such that $\langle dW_i^{\mu}(t) dW_j^\nu(t)\rangle = \delta_{ij}\delta^{\mu\nu}dt$, and $\langle dW_i^{\mu} \rangle = 0$. We can numerically solve these equations via an implicit midpoint method, aided by the use of fixed-point iteration \cite{Gardiner2004}.

For an ensemble of dynamical trajectories with initial conditions sampled from our initial Wigner distributions, quantum expectation values may then be approximated via $\langle \hat{a}(t)\rangle \approx \overline{\alpha(t)}$, $\langle \hat{\sigma}_i^\mu(t)\rangle/2 \approx \overline{s_i^\mu(t)}$, where $\overline{\,\cdot\,}$ denotes averaging with respect to this ensemble. Likewise, symmetrically-ordered correlators may be obtained via $\langle(\hat{\sigma}_i^\mu\hat{\sigma}_j^\nu + \hat{\sigma}_j^\nu\hat{\sigma}_i^\mu)(t)\rangle/8 \approx \overline{s_i^\mu(t)s_j^\nu(t)}$. Given the generic nonlinear nature of our classical equations of motion, this averaging produces results beyond mean-field theory that take into account the effect of the quantum noise distribution on our dynamics.

To benchmark this method for our squeezing protocol described in the main text, we compare against master equation calculations for small $N$. Taking $\delta = 5g_{\mathrm{ion}} = |\Omega|$ as used throughout our calculations the main text, we compute the squeezing parameter $\xi^2$ as a function of the BS evolution time $g_{\mathrm{ion}}T_{\mathrm{tr}}$ for $N=3$, $6$, and $9$. We consider select values of the dissipation rates $\gamma_+/g_{\mathrm{ion}}$ and $\gamma_z/g_{\mathrm{ion}}$ representative of the range of values considered in the main text; we also consider select TMS evolution times $g_{\mathrm{ion}}T_{\mathrm{sq}}$ up to the predicted optimal value $(\log N)/4$. For the analogous master equation calculations, we utilize a truncated Fock basis for the bosonic degree of freedom, which is chosen to ensure convergence of our dynamics.

\begin{figure}[tbh]\centering
\includegraphics[width=0.98\textwidth]{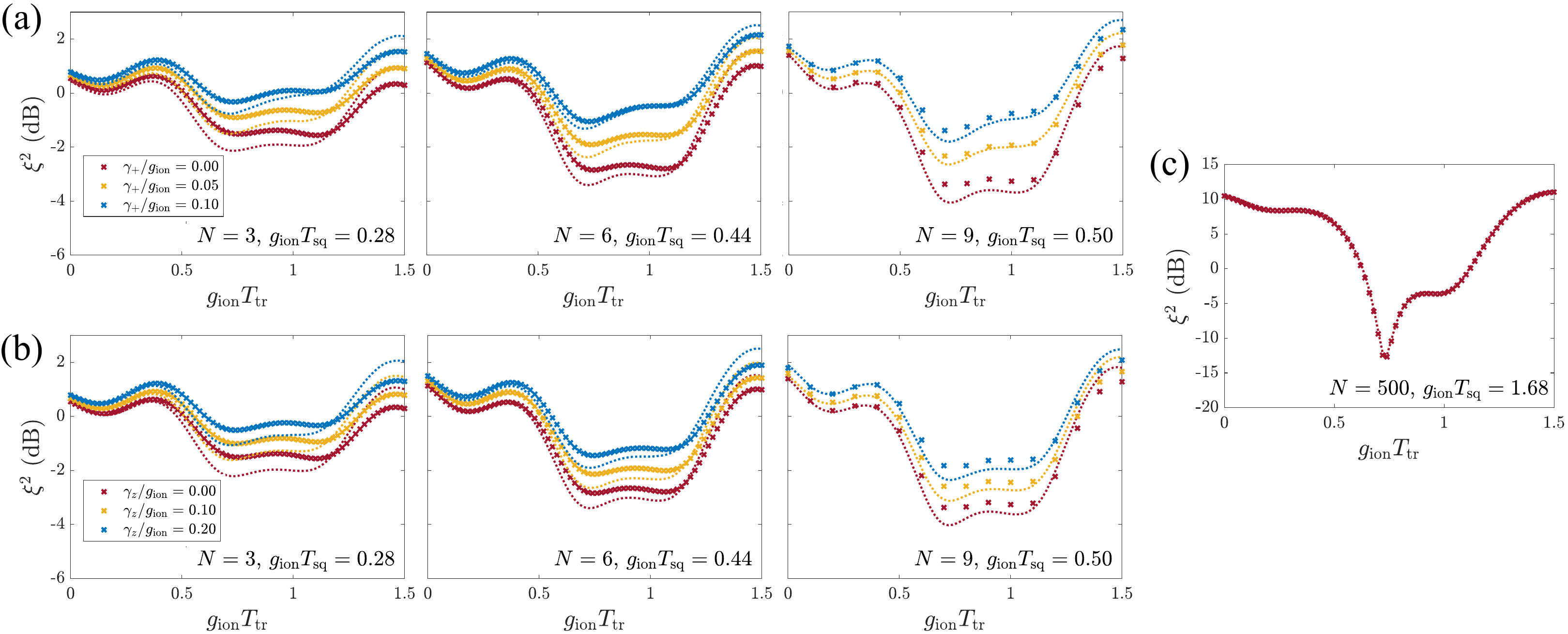}
\caption{Comparison of DDTWA solutions and exact results for the squeeze and transfer protocol. We compare the squeezing dynamics obtained via DDTWA (dotted) and solutions of the master equation (x's) for various system sizes $N$, as well as a) a range of Raman decoherence rates $\gamma_+/g_{\mathrm{ion}}$ (top) and b) Rayleigh decoherence rates $\gamma_z/g_{\mathrm{ion}}$ (bottom). We plot the spin squeezing dynamics versus $T_{\mathrm{tr}}$ for TMS evolution times $g_{\mathrm{ion}}T_{\mathrm{sq}} \approx (\log N)/4$, which is close to the theoretical optimal protocol. c) For a large system with $N=500$, we show analogous results when $\gamma_+ = \gamma_z = 0$, which allows us to compare the results of DDTWA (dotted) to our solutions of the time-dependent Schr{\"o}dinger equation (x's). We select the optimal TMS evolution time as determined by optimizing over $T_{\mathrm{sq}}$ and $T_{\mathrm{tr}}$. All results shown for $\delta = |\Omega| = 5g_{\mathrm{ion}}$. DDTWA results averaged over 10,000 trajectories in a) and b), and 1,000 trajectories in c).}
\label{fig:supp2}
\end{figure}

The results, shown in Fig.~\ref{fig:supp2}a, demonstrate that DDTWA approximately captures the resulting features of the dynamics for $\xi^2$ --- a sensitive measure of both one- and two-body observables --- over the range of considered parameters and protocol times. Our results generally appear to improve in accuracy for both larger dissipation and larger $N$, and we expect DDTWA to provide accurate solutions even for $\gamma_+ = \gamma_z = 0$ when considering the large system sizes relevant to the main text. To further verify this expectation, we compare the results of DDTWA to our solutions of the time-dependent Schr{\"o}dinger equation for the fully-coherent case of $\gamma_+ = \gamma_z = 0$ and with $N=500$ (see Fig.~\ref{fig:supp2}b), finding perfect agreement within our sampling error.

\section{Ideal sensitivity for time-reversal protocol}
In terms of $\hat{H}_{\text{TMS}}$ in Eq.~(\ref{eqn:SuppIdealTMSandBS}), the ideal time reversal protocol is mathematically defined by the following sequence of evolutions:
\begin{equation}
    \ket{\psi_\phi}=e^{iT_2\hat{H}_{\text{TMS}}}e^{-i\phi\hat{S}_z}e^{-iT_1\hat{H}_{\text{TMS}}}\ket{0_a}\otimes\ket{0}_b,
\end{equation}
where we left understood that $\hat{S}_z$ must be replaced by its bosonic approximation $\sqrt{N}(\hat{b}+\hat{b}^\dagger)/(2)$. Selecting $\hat{S}_y$ as our observable of choice, the sensitivity is defined by
\begin{equation}\label{eqn:sensdefn}
    (\delta\phi)^2=\frac{\braket{\psi_\phi|\hat{S}_y^2|\psi_\phi}}{(\partial_\phi\braket{\psi_\phi|\hat{S}_y|\psi_\phi})^2}.
\end{equation}
Though this quantity is a function of $\phi$ itself, we choose to evaluate it around $\phi=0$ since we are interested in the sensitivity to small rotations. At this point,
\begin{align}
    \begin{split}
        \braket{\psi_\phi|\hat{S}_y^2|\psi_\phi}|_{\phi=0}&\approx-\frac{N}{4}\braket{0_a,0_b|\big(e^{i(T_1-T_2)\hat{H}_{\text{TMS}}}\big)(\hat{b}-\hat{b}^\dagger)^2 \big(e^{-i(T_1-T_2)\hat{H}_{\text{TMS}}}\big)|0_a,0_b}\\[10pt]
        \partial_\phi\braket{\psi_\phi|\hat{S}_y|\psi_\phi}|_{\phi=0}&\approx\frac{N}{4}\braket{0_a,0_b|\bigg[\big(e^{i(T_1-T_2)\hat{H}_{\text{TMS}}}\big)(\hat{b}-\hat{b}^\dagger) \big(e^{-i(T_1-T_2)\hat{H}_{\text{TMS}}}\big),\big(e^{iT_1\hat{H}_{\text{TMS}}}\big)(\hat{b}+\hat{b}^\dagger) \big(e^{-iT_1\hat{H}_{\text{TMS}}}\big)\bigg]|0_a,0_b}.
    \end{split}
\end{align}
We now use the relations 
\begin{align}\begin{split}
    e^{it\hat{H}_{\text{TMS}}}\,\hat{b}\,e^{-it\hat{H}_{\text{TMS}}}&=\hat{b}\cosh(Gt)-i\hat{a}^\dagger\sinh(Gt)\\
    e^{it\hat{H}_{\text{TMS}}}\,\hat{a}\,e^{-it\hat{H}_{\text{TMS}}}&=\hat{a}\cosh(Gt)-i\hat{b}^\dagger\sinh(Gt),
\end{split}\end{align}
and find that
\begin{align}\label{eqn:Syexpecs}
    \begin{split}
        \braket{\psi_\phi|\hat{S}_y^2|\psi_\phi}|_{\phi=0}&\approx\frac{N\cosh[2G(T_1-T_2)]}{4}\\
        \partial_\phi\braket{\psi_\phi|\hat{S}_y|\psi_\phi}|_{\phi=0}&\approx\frac{N}{2}\cosh[GT_2].
    \end{split}
\end{align}
\subsection{Equal time protocol}
In the standard time-reversal protocol we set $T_1=T_2=T/2$, where $T$ is the total protocol duration time. Then plugging Eq.~(\ref{eqn:Syexpecs}) into Eq.~(\ref{eqn:sensdefn}) we obtain,
\begin{equation}
    (\delta\phi)^2=\frac{1}{N\cosh(GT/2)^2},
\end{equation}
which is Eq.~(4) of the main text. Note that the standard quantum limit is defined here as $(\delta\phi)^2_{\text{SQL}}=1/N$, so that the metrological gain is determined entirely by the $\cosh(GT/2)$ factor. At long enough times $GT\gg 1$, we can approximate the hyperbolic cosine by an exponential, so
\begin{equation}
    \frac{(\delta\phi)^2}{(\delta\phi)^2_{\text{SQL}}}=N(\delta\phi)^2=4e^{-GT}\approx e^{-(GT-1.4)}.
\end{equation}
For long times, we lose a factor of $1/2$ in the exponential time constant as compared to the original squeezing protocol and an effective time offset invariably appears, but we gain the ability to get close to the Heisenberg limit, as will be discussed in the section on exact simulations.
\subsection{Unequal reversal time}
More generally, we can optimize the sensitivity over $T_2$ given a fixed total duration $T=T_1+T_2$. This is motivated by the observation that the variance is second order insensitive at the point $T_1=T_2$ while the signal is only first order insensitive. In other words, the signal grows faster than the variance and so an improvement could be expected by moving slightly away from the equal time point. Setting $T_2=T_1+\Delta T$, $T_2+T_1=T$ and again assuming that $GT\gg 1, G\Delta T$, the sensitivity is approximately
\begin{equation}
    N(\delta\phi)^2\approx 4 e^{-GT}e^{-G\Delta T}\cosh(2G\Delta T)=2 e^{-GT}(e^{-3G\Delta T}+e^{G\Delta T}).
\end{equation}
Optimizing over $\Delta T$ we find that 
\begin{equation}
    e^{4G\Delta T_{\text{opt}}}=3,
\end{equation}
and so 
\begin{equation}
    N(\delta\phi)^2_{\text{opt}}\approx 3.5 e^{-GT}.
\end{equation}
This is a gain of about $0.5$dB compared to the case with $T_1 = T_2$. However, as discussed in Sec.~\ref{sec:ExactNumerics}, in the context of exact numerical calculations, the overall gain when incorporating expected finite size effects is actually about $3$dB. In other words, the over-reversal seems to partially paliate the effects of finite $N$, allowing the increase of the metrological gain to be sustained for a longer time.

\begin{figure}
    \centering
    \includegraphics[width=0.64\textwidth]{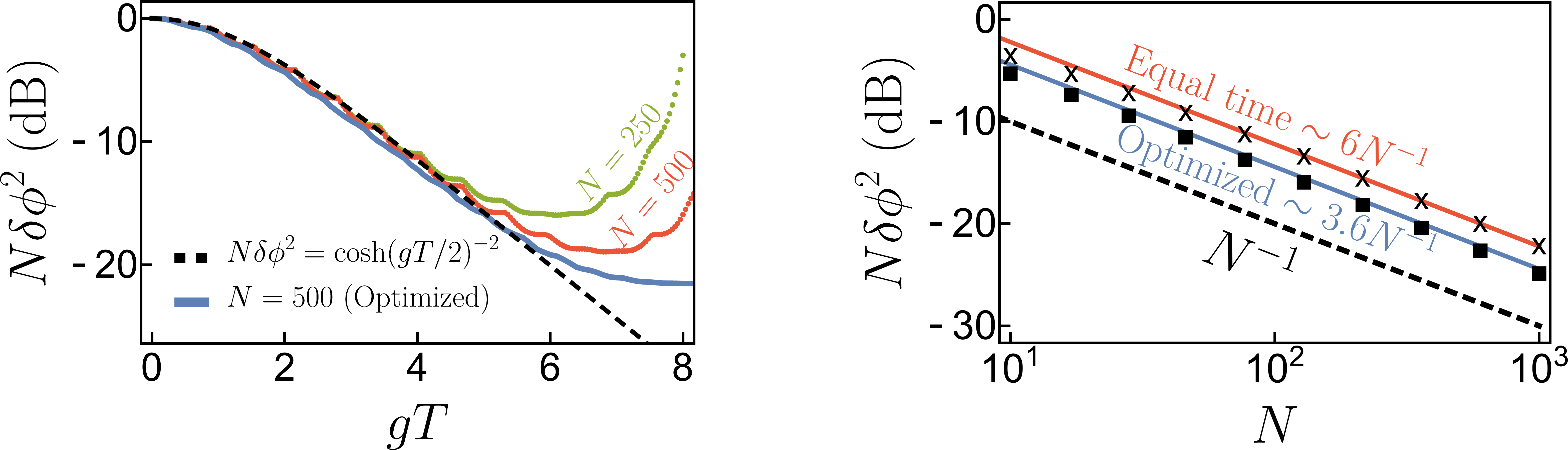}
    \includegraphics[width=0.35\textwidth]{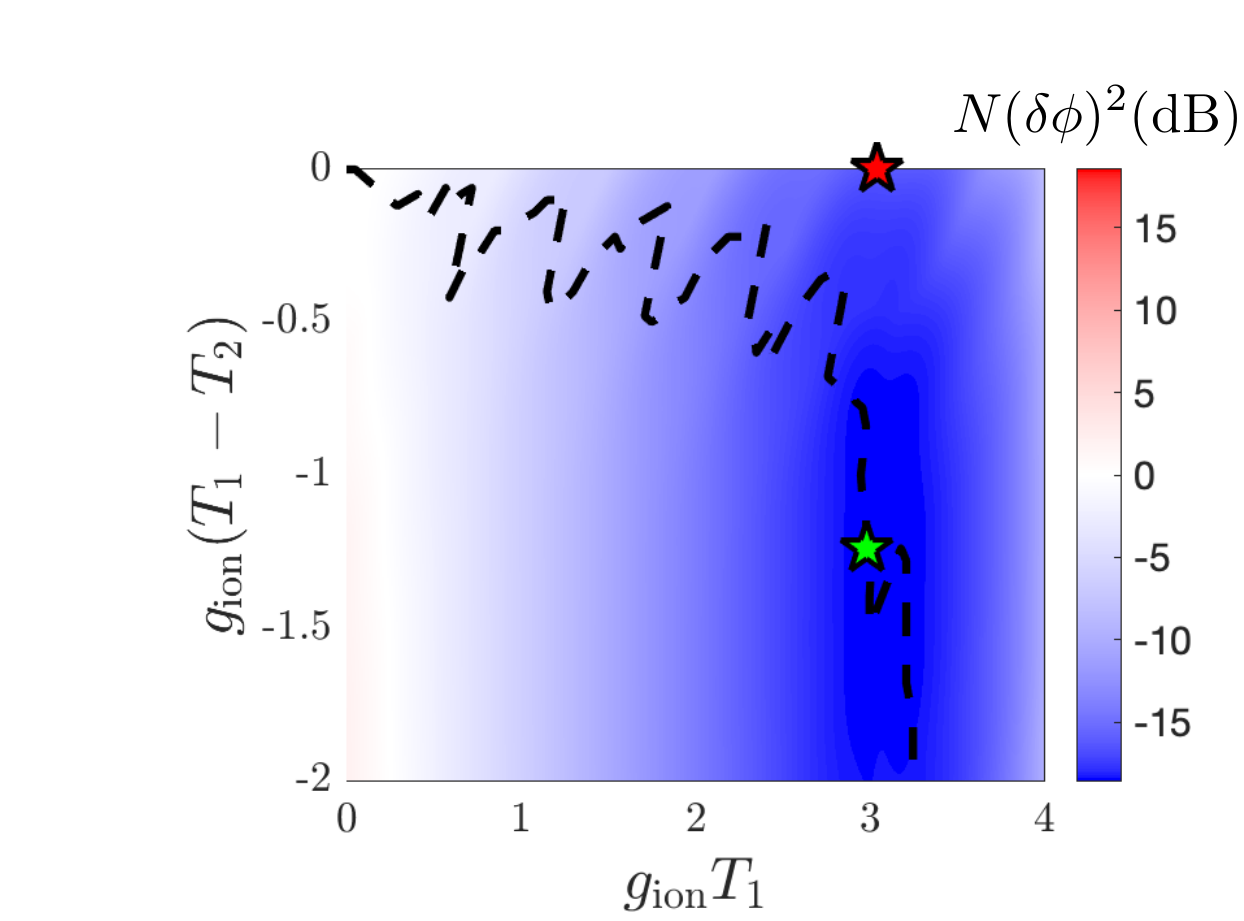}
    \caption{(Left) Metrological gain as a function of total protocol time $T=T_1+T_2$ for the ion implementation. We show the ideal equal time $T_1=T_2$, $N,\delta\to\infty$ case (dashed black line), and the results for finite $\delta=5g_{\Rev{\text{ion}}}$ (see main text) and $N=250$ (green dots), $N=500$ (red dots). We also include the sensitivity for an optimized version of the protocol where $T_2$ is slightly larger than $T_1$ and saturates closer to the Heisenberg limit (blue line). (Center) Scaling of the optimal sensitivity for both schemes as a function of $N$. (Right) Sensitivity for a range of forward and backward evolution times for $N=250$ and $\delta = 5g_{\mathrm{ion}}$. The dashed line indicates the minimum sensitivity when optimized over fixed protocol times $T = T_1 + T_2$. The green star indicates the best optimized protocol with $g_{\mathrm{ion}}T_{1} = 2.98$ and $g_{\mathrm{ion}}(T_1 - T_2) = -1.24$, yielding a metrological gain of $18.5$ dB. The red star indicates the best equal time protocol ($T_1 = T_2$), with $g_{\mathrm{ion}}T_{\mathrm{sq}} = 3.04$ and yielding a metrological gain of $16.3$ dB.}
    \label{fig:SupptimeReversal}
\end{figure}

We show exact simulations of the time-reversal protocol for the ion implementation including finite $N$ and $\delta$ in Fig.~(\ref{fig:SupptimeReversal}) (left), including also the the unequal reversal time scheme described in the previous section. In the right panel we show the scaling of the optimal sensitivity with $N$ for both equal and unequal reversal times, and demonstrate that they attaing $1/N$ scaling. Exact numerical simulations are done using the methods described in section~\ref{sec:ExactNumerics} of this Supplementary Material.

\section{Twist-and-turn}
\begin{figure}
    \centering
    \includegraphics[width=0.98\textwidth]{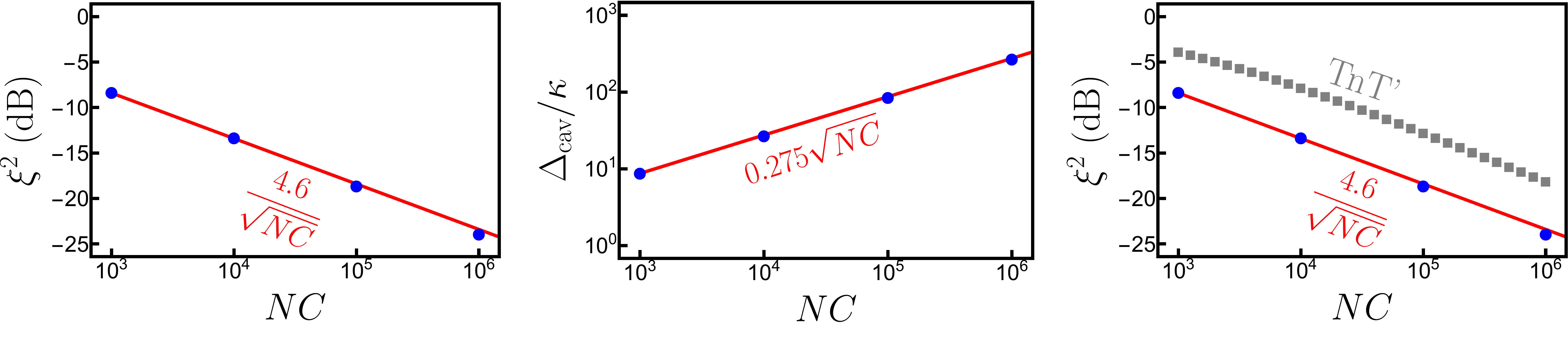}
    \caption{Twist-and-turn. (Left) Optimal squeezing as a function of $NC$ obtained by numerically solving Eq.~\ref{eqn:SuppTnTEquations} (blue dots) compared to the theoretical approximation $\xi^2_{\text{opt}}=4.6/\sqrt{NC}$ (solid red line). (Center) Optimal detuning as a function of $NC$. (Right) Optimal squeezing when constrained to $\Delta_{\text{cav}}\geq 10 g_{\text{cav}}\sqrt{N}$ (gray squares) compared to unconstrained twist-and-turn (blue dots, red line).}
    \label{fig:SuppTnT}
\end{figure}

Here we provide a treatment of the twist-and-turn protocol in the large $N$ limit in the implementation of Ref.~\cite{Hu_2017}, which includes both single particle and collective emission (the latter coming from photons leaking from the cavity). While Ref.~\cite{Hu_2017} provides the scaling with particle number and cavity cooperativity, the prefactor is missing, so we determine it here. The evolution equation of the system is
\begin{equation}\label{eqn:SuppTwistAndTurn}
    \partial_t\hat{\rho}=-i\Big[\chi\hat{S}^+\hat{S}^--\frac{\chi N}{2}\hat{S}_x,\hat{\rho}\Big]+\Gamma\Big(\hat{S}^-\hat{\rho}\hat{S}^+-\frac{1}{2}\{\hat{S}^+\hat{S}^-,\hat{\rho}\}\Big)+\gamma\sum_{i=1}^N\Big(\hat{\sigma}^-_i\hat{\rho}\hat{\sigma}_i^+-\frac{1}{2}\{\hat{\sigma}_i^+\hat{\sigma}_i^-,\hat{\rho}\}\Big),
\end{equation}
where $\chi=4g_{\text{cav}}^2\Delta_{\text{cav}}/(4\Delta_{\text{cav}}^2+\kappa^2)$ and $\Gamma/\chi=\kappa/\Delta_{\text{cav}}$. We proceed as in Supplementary Section~\ref{sec:SuppSpontaneousEmission} by first calculating the mean field equations of motion associated to $\hat{S}^-$ and $\hat{S}_z$:
\begin{align}
    \begin{split}
        \dot{s}_{\text{mf}}&=i\frac{\chi N}{2}z_{\text{mf}}+(\Gamma-2i\chi) s_{\text{mf}}z_{\text{mf}}-\gamma s_{\text{mf}}/2\\
        \dot{z}_{\text{mf}}&=-\frac{\chi N}{2}\frac{(s_{\text{mf}}^*-s_{\text{mf}})}{2}-\Gamma s_{\text{mf}}^*s_{\text{mf}}-\gamma\Big(z_{\text{mf}}+\frac{N}{2}\Big),
    \end{split}
\end{align}
with initial conditions $z_{\text{mf}}=0$ and $s_{\text{mf}}=N/2$, and define the mean field vector $v=\big[\text{Re}(s_{\text{mf}}),\text{Im}(s_{\text{mf}}),z_{\text{mf}}\big]$. We then calculate the equations for the second moments of the fluctuations $(\delta \hat{S}^+,\delta\hat{S}_z)\equiv (\hat{S}^+-s_{\text{mf}}^*,\,\hat{S}_z-z_{\text{mf}})$:
% \begin{align}
%     \begin{split}
%         \dot{F}_{xx}&=2F_{xz}\text{Re}\Big[\big(\Gamma-2i\chi\big)s_{\text{mf}}^*\Big]+\big(2\Gamma z_{\text{mf}}-\gamma\big)F_{xx}+4\chi F_{xy}+\frac{\gamma N}{2}+2\Gamma z_{\text{mf}}^2\\
%         \dot{F}_{yy}&=2F_{yz}\text{Im}\Big[\big(\Gamma-2i\chi\big)s_{\text{mf}}^*\Big]-\chi N F_{yz}+\big(2\Gamma z_{\text{mf}}-\gamma\big)F_{yy}-4\chi F_{xy}+\frac{\gamma N}{2}+2\Gamma z_{\text{mf}}^2\\
%         \dot{F}_{xy}&=F_{xz}\text{Im}\Big[\big(\Gamma-2i\chi\big)s_{\text{mf}}^*\Big]-F_{yz}\text{Re}\Big[\big(\Gamma-2i\chi\big)s_{\text{mf}}^*\Big]+\frac{\chi N}{2}F_{xz}+(2\Gamma z_{\text{mf}}-\gamma)F_{xy}-2\chi (F_{xx}-F_{yy})
%     \end{split}
% \end{align}
\begin{align}\begin{split}\label{eqn:SuppTnTEquations}
    \partial_t\braket{\delta\hat{S}^+\delta\hat{S}^+}&=\bigg[\big(\Gamma-2i\chi\big)s_{\text{mf}}^*+i\frac{\chi N}{2}\bigg]\Big[2\braket{\delta\hat{S}^+\delta\hat{S}_z}+s_{\text{mf}}\Big]+\Big[2\big(\Gamma-2i\chi\big)z_{\text{mf}}-\gamma\Big]\braket{\delta\hat{S}^+\delta\hat{S}^+}\\[7pt]
    \partial_t\braket{\delta\hat{S}^+\delta\hat{S}^-}&=2\text{ Re}\Bigg\{\bigg[\big(\Gamma+2i\chi\big)s_{\text{mf}}-i\frac{\chi N}{2}\bigg]\braket{\delta\hat{S}^+\delta\hat{S}_z}\Bigg\}+\big(2\Gamma z_{\text{mf}}-\gamma\big)\braket{\delta\hat{S}^+\delta\hat{S}^-}\\[7pt]
    \partial_t\braket{\delta\hat{S}^+\delta\hat{S}_z}&=\bigg[\big(\Gamma-2i\chi\big)s_{\text{mf}}^*+i\frac{\chi N}{2}\bigg]\braket{\delta\hat{S}_z^2}+\bigg[\big(\Gamma-2i\chi\big)z_{\text{mf}}-\frac{3\gamma}{2}\bigg]\braket{\delta\hat{S}^+\delta\hat{S}_z}\\[5pt]
    &\hspace{4.90cm}+\bigg(\frac{i\chi N}{4}-\Gamma s_{\text{mf}}\bigg)\braket{\delta\hat{S}^+\delta\hat{S}^+}-\bigg(\Gamma s_{\text{mf}}^*+\frac{i\chi N}{4}\bigg)\braket{\delta\hat{S}^+\delta\hat{S}^-}\\[7pt]
    \partial_t\braket{\delta\hat{S}_z^2}&=-\chi N\text{ Im}\big[\braket{\delta\hat{S}^+\delta\hat{S}_z}+s_{\text{mf}}^*/2\big]-4\Gamma\text{ Re}\Big[\braket{\delta\hat{S}^+\delta\hat{S}_z}s_{\text{mf}}^*\Big]-2\gamma \braket{\delta\hat{S}^2_z}+\gamma\bigg(z_{\text{mf}}+\frac{N}{2}\bigg)-\Gamma s_{\text{mf}}^* s_{\text{mf}}
\end{split}\end{align}
with initial conditions $\braket{\delta\hat{S}^+\delta\hat{S}^+}=-N/4$, $\braket{\delta\hat{S}^+\delta\hat{S}^-}=N/4$, $\braket{\delta\hat{S}^+\delta\hat{S}_z}=-N/4$ and $\braket{\delta\hat{S}_z^2}=N/4$. We have neglected terms in the equation of motion with $\braket{\delta\hat{S}^3_{+,z}}$ since they are a factor $\sqrt{N}$ smaller than the terms kept in the previous equations. From these 4 complex correlators we can obtain the covariance matrix $M_{ij}=\braket{\{\delta\hat{S}_i,\delta\hat{S}_j\}}/2$, for $i,j=x,y,z$, with
\begin{align}\begin{split}
    M_{xx}&=\frac{1}{2}\Big[\braket{\delta\hat{S}^+\delta\hat{S}^-}+\text{Re}\big[\braket{\delta\hat{S}^+\delta\hat{S}^+}\big]-z_{\text{mf}}\Big]\hspace{1cm}M_{zx}=\text{Re}\Big[\braket{\delta\hat{S}^+\delta\hat{S}_z}+s_{\text{mf}}^*\Big]\hspace{1cm} M_{xy}=\text{Im}\big[\braket{\delta\hat{S}^+\delta\hat{S}^+}\big]/2\\
    M_{yy}&=\frac{1}{2}\Big[\braket{\delta\hat{S}^+\delta\hat{S}^-}-\text{Re}\big[\braket{\delta\hat{S}^+\delta\hat{S}^+}\big]-z_{\text{mf}}\Big]\hspace{1cm}M_{zy}=\text{Im}\Big[\braket{\delta\hat{S}^+\delta\hat{S}_z}+s_{\text{mf}}^*\Big].
\end{split}\end{align}
To calculate the minimal variance, we project this matrix onto the subspace orthogonal to $v$: $M_p=(1-n n^T)M(1-nn^T)$, where $n=v/|v|$ is the normalized mean field vector and compute its smallest nonzero eigenvalue. Squeezing (incorporating contrast decay) is calculated by dividing this minimal variance by $|v|^2/N$. The time-dependence of $z_{\text{mf}}$ and $s_{\text{mf}}$ prevents us from obtaining a closed-form solution to the equations for fluctuations, so we resort to numerical simulation. We define a dimensionless time $\tau=\chi N t$ and parameterize \begin{equation}
    \frac{\Gamma}{\chi}=\frac{\kappa}{\Delta_{\text{cav}}}\equiv \eta\hspace{1cm}\frac{\gamma}{\chi N}=\frac{1}{NC}\bigg(\eta+\frac{4}{\eta}\bigg)
\end{equation}
in terms of $\eta$ (the inverse detuning, measured in units of $\kappa$) and the collective cooperativity $NC=4g_{\text{cav}}^2N/(\kappa\gamma)$. For different values of $NC$, we optimize over evolution time and $\eta$ to obtain the best possible squeezing. The results are shown in Fig.~\ref{fig:SuppTnT} (left and center, blue dots) together with the theoretical approximations $\xi^2_{\text{opt}}\approx 4.57/\sqrt{NC}$ and $\Delta_{\text{cav}}^{\text{opt}}/\kappa\approx 0.275\sqrt{NC}$ (red lines) which correctly capture the dependence with $NC$. 

If we use the estimate for the optimal detuning, we find that $\Delta_{\text{cav}}^{\text{opt}}/(g_{\text{cav}}\sqrt{N})\approx 0.55\sqrt{\kappa/\gamma}$. For the cavity parameters used in the paper [$(\gamma,\kappa)=2\pi (7.5,153)$ kHz], this corresponds to $\Delta_{\text{cav}}^{\text{opt}}/(g_{\text{cav}}\sqrt{N})\approx 2.5$, which is outside the regime of validity of Eq.~(\ref{eqn:SuppTwistAndTurn}). To remedy this, we also numerically optimize squeezing over time at a larger fixed detuning $\Delta_{\text{cav}}=1.1\sqrt{NC}$, which guarantees $\Delta_{\text{cav}}/(g_{\text{cav}}\sqrt{N})\approx 10$ (for the same values of $\kappa$ and $\gamma$), and gives the gray lines labeled TnT' in Fig.~3 in the main text and in Fig.~\ref{fig:SuppTnT} (right).

\bibliography{library}